\documentclass[prl,superscriptaddress,lineno]{revtex4-1}
\usepackage{graphicx}
\usepackage{dcolumn}
\usepackage{bm}

\begin{document}

\title{Measurement of neutrino-induced neutral-current coherent $\pi^0$ production in the NOvA near detector}

%
\newcommand{\ANL}{Argonne National Laboratory, Argonne, Illinois 60439, 
USA}
\newcommand{\ICS}{Institute of Computer Science, The Czech 
Academy of Sciences, 
182 07 Prague, Czech Republic}
\newcommand{\IOP}{Institute of Physics, The Czech 
Academy of Sciences, 
182 21 Prague, Czech Republic}
\newcommand{\Atlantico}{Universidad del Atlantico,
Km. 7 antigua via a Puerto Colombia, Barranquilla, Colombia}
\newcommand{\BHU}{Department of Physics, Institute of Science, Banaras 
Hindu University, Varanasi, 221 005, India}
\newcommand{\UCLA}{Physics and Astronomy Department, UCLA, Box 951547, Los 
Angeles, California 90095-1547, USA}
\newcommand{\Caltech}{California Institute of 
Technology, Pasadena, California 91125, USA}
\newcommand{\Cochin}{Department of Physics, Cochin University
of Science and Technology, Kochi 682 022, India}
\newcommand{\Charles}
{Charles University, Faculty of Mathematics and Physics,
 Institute of Particle and Nuclear Physics, Prague, Czech Republic}
\newcommand{\Cincinnati}{Department of Physics, University of Cincinnati, 
Cincinnati, Ohio 45221, USA}
\newcommand{\CSU}{Department of Physics, Colorado 
State University, Fort Collins, CO 80523-1875, USA}
\newcommand{\CTU}{Czech Technical University in Prague,
Brehova 7, 115 19 Prague 1, Czech Republic}
\newcommand{\Dallas}{Physics Department, University of Texas at Dallas,
800 W. Campbell Rd. Richardson, Texas 75083-0688, USA}
\newcommand{\DallasU}{University of Dallas, 1845 E 
Northgate Drive, Irving, Texas 75062 USA}
\newcommand{\Delhi}{Department of Physics and Astrophysics, University of 
Delhi, Delhi 110007, India}
\newcommand{\JINR}{Joint Institute for Nuclear Research,  
Dubna, Moscow region 141980, Russia}
\newcommand{\FNAL}{Fermi National Accelerator Laboratory, Batavia, 
Illinois 60510, USA}
\newcommand{\UFG}{Instituto de F\'{i}sica, Universidade Federal de 
Goi\'{a}s, Goi\^{a}nia, Goi\'{a}s, 74690-900, Brazil}
\newcommand{\Guwahati}{Department of Physics, IIT Guwahati, Guwahati, 781 
039, India}
\newcommand{\Harvard}{Department of Physics, Harvard University, 
Cambridge, Massachusetts 02138, USA}
\newcommand{\Houston}{Department of Physics, 
University of Houston, Houston, Texas 77204, USA}
\newcommand{\IHyderabad}{Department of Physics, IIT Hyderabad, Hyderabad, 
502 205, India}
\newcommand{\Hyderabad}{School of Physics, University of Hyderabad, 
Hyderabad, 500 046, India}
\newcommand{\IIT}{Department of Physics,
Illinois Institute of Technology,
Chicago IL 60616, USA}
\newcommand{\Indiana}{Indiana University, Bloomington, Indiana 47405, 
USA}
\newcommand{\INR}{Inst. for Nuclear Research of Russia, Academy of 
Sciences 7a, 60th October Anniversary prospect, Moscow 117312, Russia}
\newcommand{\Iowa}{Department of Physics and Astronomy, Iowa State 
University, Ames, Iowa 50011, USA}
\newcommand{\Irvine}{Department of Physics and Astronomy, 
University of California at Irvine, Irvine, California 92697, USA}
\newcommand{\Jammu}{Department of Physics and Electronics, University of 
Jammu, Jammu Tawi, 180 006, Jammu and Kashmir, India}
\newcommand{\Lebedev}{Nuclear Physics and Astrophysics Division, Lebedev 
Physical 
Institute, Leninsky Prospect 53, 119991 Moscow, Russia}
\newcommand{\MSU}{Department of Physics and Astronomy, Michigan State 
University, East Lansing, Michigan 48824, USA}
\newcommand{\Duluth}{Department of Physics and Astronomy, 
University of Minnesota Duluth, Duluth, Minnesota 55812, USA}
\newcommand{\Minnesota}{School of Physics and Astronomy, University of 
Minnesota Twin Cities, Minneapolis, Minnesota 55455, USA}
\newcommand{\Oxford}{Subdepartment of Particle Physics, 
University of Oxford, Oxford OX1 3RH, United Kingdom}
\newcommand{\Panjab}{Department of Physics, Panjab University, 
Chandigarh, 160 014, India}
\newcommand{\Pitt}{Department of Physics, 
University of Pittsburgh, Pittsburgh, Pennsylvania 15260, USA}
\newcommand{\RAL}{Rutherford Appleton Laboratory, Science and 
Technology Facilities Council, Didcot, OX11 0QX, United Kingdom}
\newcommand{\SAlabama}{Department of Physics, University of 
South Alabama, Mobile, Alabama 36688, USA} 
\newcommand{\Carolina}{Department of Physics and Astronomy, University of 
South Carolina, Columbia, South Carolina 29208, USA}
\newcommand{\SDakota}{South Dakota School of Mines and Technology, Rapid 
City, South Dakota 57701, USA}
\newcommand{\SMU}{Department of Physics, Southern Methodist University, 
Dallas, Texas 75275, USA}
\newcommand{\Stanford}{Department of Physics, Stanford University, 
Stanford, California 94305, USA}
\newcommand{\Sussex}{Department of Physics and Astronomy, University of 
Sussex, Falmer, Brighton BN1 9QH, United Kingdom}
\newcommand{\Syracuse}{Department of Physics, Syracuse University,
Syracuse NY 13210, USA}
\newcommand{\Tennessee}{Department of Physics and Astronomy, 
University of Tennessee, Knoxville, Tennessee 37996, USA}
\newcommand{\Texas}{Department of Physics, University of Texas at Austin, 
Austin, Texas 78712, USA}
\newcommand{\Tufts}{Department of Physics and Astronomy, Tufts University, Medford, 
Massachusetts 02155, USA}
\newcommand{\UCL}{Physics and Astronomy Dept., University College London, 
Gower Street, London WC1E 6BT, United Kingdom}
\newcommand{\Virginia}{Department of Physics, University of Virginia, 
Charlottesville, Virginia 22904, USA}
\newcommand{\WSU}{Department of Mathematics, Statistics, and Physics,
 Wichita State University, 
Wichita, Kansas 67206, USA}
\newcommand{\WandM}{Department of Physics, William \& Mary, 
Williamsburg, Virginia 23187, USA}
\newcommand{\Winona}{Department of Physics, Winona State University, P.O. 
Box 5838, Winona, Minnesota 55987, USA}
\newcommand{\Wisconsin}{Department of Physics, University of 
Wisconsin-Madison, Madison, Wisconsin 53706, USA}
\newcommand{\Crookston}{Math, Science and Technology Department, University 
of Minnesota -- Crookston, Crookston, Minnesota 56716, USA}
\newcommand{\deceased}{Deceased.}
\affiliation{\ANL}
\affiliation{\Atlantico}
\affiliation{\BHU}
\affiliation{\Caltech}
\affiliation{\Charles}
\affiliation{\Cincinnati}
\affiliation{\Cochin}
\affiliation{\CSU}
\affiliation{\CTU}
\affiliation{\DallasU}
\affiliation{\Delhi}
\affiliation{\FNAL}
\affiliation{\UFG}
\affiliation{\Guwahati}
\affiliation{\Harvard}
\affiliation{\Houston}
\affiliation{\Hyderabad}
\affiliation{\IHyderabad}
\affiliation{\Indiana}
\affiliation{\ICS}
\affiliation{\IIT}
\affiliation{\INR}
\affiliation{\IOP}
\affiliation{\Iowa}
\affiliation{\Irvine}
\affiliation{\Jammu}
\affiliation{\JINR}
\affiliation{\Lebedev}
\affiliation{\MSU}
\affiliation{\Duluth}
\affiliation{\Minnesota}
\affiliation{\Panjab}
\affiliation{\Pitt}
\affiliation{\SAlabama}
\affiliation{\Carolina}
\affiliation{\SDakota}
\affiliation{\SMU}
\affiliation{\Stanford}
\affiliation{\Sussex}
\affiliation{\Syracuse}
\affiliation{\Tennessee}
\affiliation{\Texas}
\affiliation{\Tufts}
\affiliation{\UCL}
\affiliation{\Virginia}
\affiliation{\WSU}
\affiliation{\WandM}
\affiliation{\Wisconsin}

\author{M.~A.~Acero}
\affiliation{\Atlantico}

\author{P.~Adamson}
\affiliation{\FNAL}


\author{L.~Aliaga}
\affiliation{\FNAL}

\author{T.~Alion}
\affiliation{\Sussex}

\author{V.~Allakhverdian}
\affiliation{\JINR}



\author{N.~Anfimov}
\affiliation{\JINR}


\author{A.~Antoshkin}
\affiliation{\JINR}

\affiliation{\Minnesota}

\author{E.~Arrieta-Diaz}
\affiliation{\SMU}


\author{A.~Aurisano}
\affiliation{\Cincinnati}


\author{A.~Back}
\affiliation{\Iowa}

% ...

\collaboration{The NOvA Collaboration}
\noaffiliation





\begin{abstract}

The cross section of neutrino-induced neutral-current coherent $\pi^0$ production on a carbon-dominated target is measured in the NOvA near detector. 
This measurement uses a narrow-band neutrino beam with an average neutrino energy of 2.7\,GeV, which is of interest to ongoing and future long-baseline neutrino oscillation experiments. 
The measured, flux-averaged cross section is $\sigma = 13.8\pm0.9 (\text{stat})\pm2.3 (\text{syst}) \times 10^{-40}\,\text{cm}^2/\text{nucleus}$, consistent with model prediction.                                  
This result is the most precise measurement of neutral-current coherent $\pi^0$ production in the few-GeV neutrino energy region. 
\end{abstract}

\maketitle

\twocolumngrid
Neutrinos can interact coherently with target nuclei and produce outgoing pions via either neutral-current (NC) or charged-current (CC) interactions. 
In the case of an NC interaction, a $\pi^0$ is produced: 
\begin{equation}
\nu {\cal A} \rightarrow \nu {\cal A} \pi^0.
\end{equation}
Coherent interactions are characterized by very small momentum transfer to the target nucleus 
with no exchange of quantum numbers, while the target nucleus remains in its ground state. 
The characteristic signal topology of NC coherent $\pi^0$ production is a single, forward-going $\pi^0$, with no other hadrons in the final state. 

There are two major motivations for measuring the NC coherent $\pi^0$ cross section. 
First, coherent $\pi^0$ production is a contribution to the background of long-baseline $\nu_{\mu}\rightarrow\nu_e$ oscillation measurements. 
In some neutrino detectors, the photons from $\pi^0$ decay are reconstructed as electromagnetic showers which are often difficult to separate from the showers induced by electrons. 
An NC $\pi^0$ event can be misidentified as a $\nu_e$-CC signal event if the two photon showers are not spatially separated, or if one shower is undetected.  
Knowledge of coherent $\pi^0$ production provides a constraint on the size of this background. 
Second, coherent pion production provides insight into the weak hadronic current structure and serves as a test of the partially conserved axial current (PCAC) hypothesis \cite{Adler:1964pcac}. 
Models based upon PCAC relate the coherent pion production to the pion-nucleus elastic scattering cross section at the $Q^2=0$ limit, and extrapolate to low but nonzero $Q^2$ values. 
Such models include the Rein-Sehgal model \cite{Rein:1983rs,Rein:2007rs} in the GENIE neutrino generator \cite{Andreopoulos:2010genie} used for this analysis. 
Further improvement of PCAC models was made by Berger \textit{et al.} (Berger-Sehgal model) \cite{Berger:2009bs} and others
\cite{Belkov:1987bk,Kopeliovich:2005bzh,Hernandez:2009eh,Kartavtsev:2006eap,Paschos:2009eap}.  
The PCAC models are known to perform well in their intended multi-GeV energy ranges. 
Their performance in NC interactions at neutrino energies of a few GeV, however, remains to be established.  
There is another class of models, referred to as microscopic models, that do not rely on PCAC. 
These models are built using pion production amplitudes at the nucleon level \cite{Singh:2006microscopic1,Alvarez-Ruso:2007microscopic2,Alvarez-Ruso:2007microscopic2nc,Amaro:2009microscopic3,Leitner:2009microscopic4,Hernandez:2009microscopic5,Nakamura:2009microscopic6}  
and are expected to be more reliable than PCAC-based models at neutrino energies below 1\,GeV, 
where the $\Delta$ resonance dominates the weak pion production.  
They typically miss contributions from higher resonances above $\Delta$ at higher neutrino energy. 

The NC coherent $\pi^0$ cross section contributes roughly 1\% of the inclusive neutrino interactions in the few-GeV neutrino energy region, much smaller than other $\pi^0$ production modes. 
This challenging situation requires the extraction of a small signal with large backgrounds. 
The backgrounds arise mainly from NC-induced baryon resonance (RES) interactions and $\pi^0$ production from NC deep-inelastic scattering (DIS) interactions, where only a single $\pi^0$ is  reconstructed. 
Diffractive (DFR) $\pi^0$ production, where neutrinos scatter off free protons (hydrogen) with small momentum transfer and produce $\pi^0$'s, also contributes to the background. 
Unlike coherent interactions, a recoil proton is sometimes visible in DFR $\pi^0$ production, which makes the DFR event topology potentially different from the coherent signal. 
The visibility of the recoil proton in a detector depends upon the proton's kinetic energy ($T_p$), which is determined by $T_p = |t|/2m_p$, where $|t|$ is the square of four-momentum transfer to the nucleus, and $m_p$ is the proton mass. 

The coherent process is best identified by a low value of $|t|$ ($|t| \lesssim\hbar^2/R^2$, where $R$ is the nuclear radius).
However, in NC interactions $|t|$ cannot be determined, because the outgoing neutrino momentum cannot be measured. 
Alternatively, distinct characteristics of the coherent process can be used to separate coherent from background $\pi^0$ production. 
First, the coherent $\pi^0$ production has no other particles in the final state and little vertex activity, 
while background processes often produce additional nucleons or pions and have larger energy depositions near the neutrino interaction vertex. 
Second, the $\pi^0$s from coherent production are distinctly forward going.  
A region with enhanced coherent signal in the 2D space of reconstructed $\pi^0$ energy and scattering angle can be defined. 
The coherent signal is measured as an excess of data events over the background prediction in this region.

There are relatively few existing NC coherent $\pi^0$ measurements. 
Early bubble chamber results suffer from large statistical uncertainties \cite{Faissner:1983ap,Isiksal:1984Gargamelle,Baltay:1986bc,Bergsma:1985charm,Grabosch:1986SKAT}. 
More recently, NOMAD, MiniBooNE, SciBooNE and MINOS reported coherent $\pi^0$ measurements with higher statistics but with systematic-limited precision \cite{Aguilar-Arevalo:2008miniboone,Kurimoto:2008sciboone,Kullenberg:2009nomad,Adamson:2016minos}.  
In particular, the measurements in the few-GeV neutrino energy region, relevant for the next-generation neutrino oscillation experiments,  
have large uncertainties.
\begin{figure*}[ht]
\begin{center}
\includegraphics[width=0.5\linewidth]{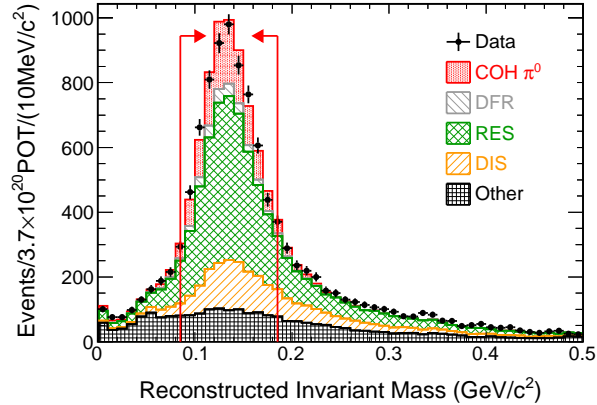}
\end{center}
\caption{Data and simulated $\pi^0$ invariant mass distribution of the selected two-prong NC $\pi^0$ sample. Data are shown as solid circles with statistical error bars. The shaded histograms represent the simulated prediction divided by interaction modes, including coherent signal and NC RES, DIS and DFR background $\pi^0$ productions. Charged current $\pi^0$ production, external events, and interactions without final-state $\pi^0$s are classified under ``other''.  Vertical lines with arrows show the range of invariant masses accepted into the analysis. 
\label{figure_mggnc}}
\end{figure*}

\begin{figure*}[ht]
\begin{center}
\includegraphics[width=0.49\linewidth]{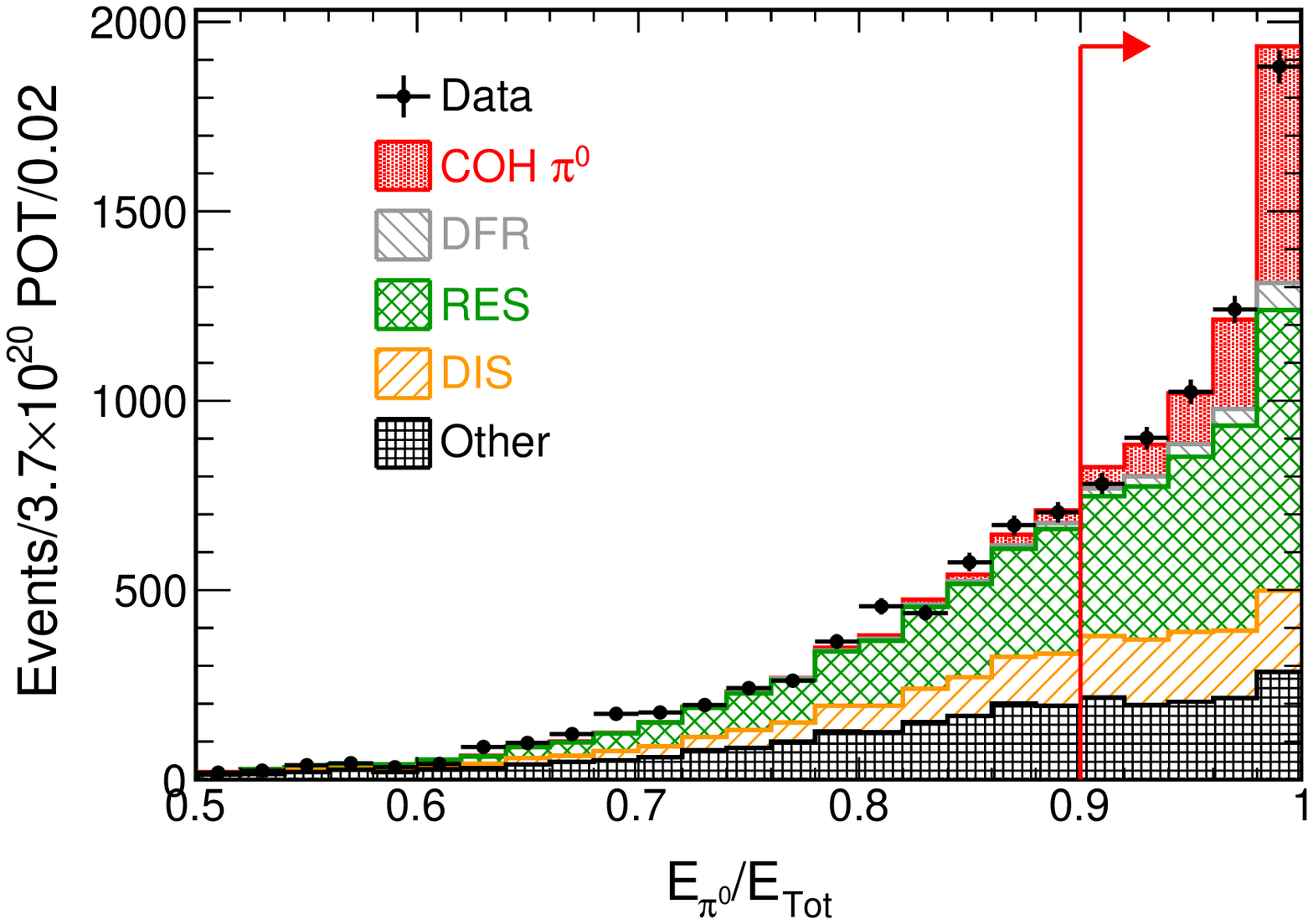}
\includegraphics[width=0.49\linewidth]{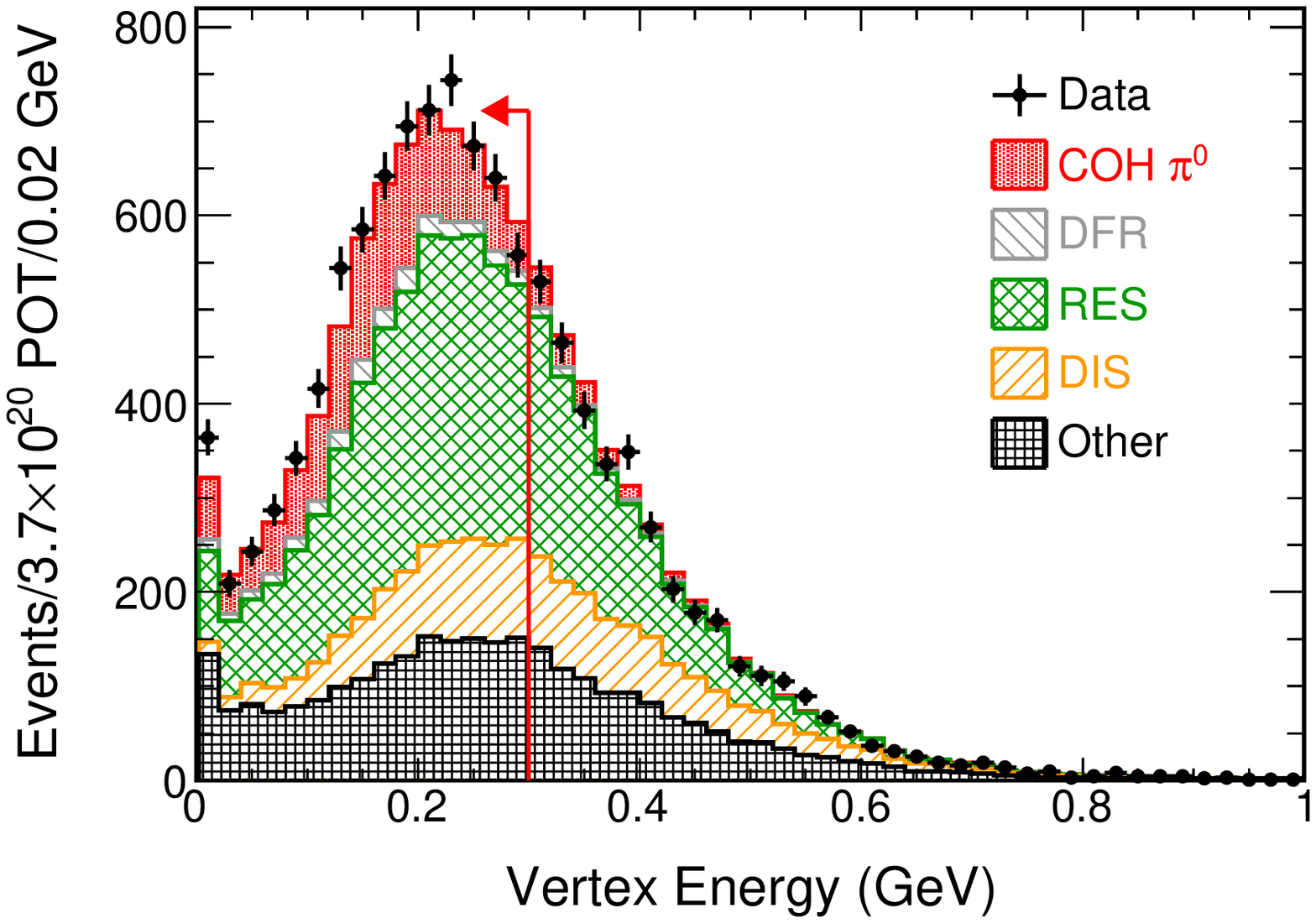}
\end{center}
\caption{
Fraction of event energy contained in the reconstructed $\pi^0$ (left) and vertex energy (right) in data (black circle) and simulation (shaded histograms). 
Statistical error bars are shown for data. The simulated distribution is classified by interaction modes. 
Events to the right (left) of the vertical red line are selected into the signal sample, and the rest of the events are selected into the control sample. 
The cut values are optimized by maximizing figure of merit ($FOM = s/\sqrt{s+ b}$, where $s$ and $b$ are the numbers of signal and background events passing the cuts).
\label{figure_caleratio_vtxe}}																																			
\end{figure*}

This paper reports a measurement of NC coherent $\pi^0$ cross section using the fine-grained sampling of neutrino interactions in a predominantly carbon tracking medium afforded by the NOvA near detector (ND) \cite{Ayres:2007nova} exposed to the off-axis flux of the NuMI beam \cite{Adamson:2016numi} at Fermilab. 
The flux-averaged cross section is defined as
\begin{equation}
\sigma = \frac{N_{\rm data} - N_{\rm bkg}}{\epsilon\times N_{\rm target} \times \phi} ,
\end{equation}
where $N_{\rm data} $ and $N_{\rm bkg}$ are the number of selected data and simulation-predicted background events, respectively,  
 $\epsilon$ is the efficiency of the coherent signal selection calculated from simulation, 
 $N_{\rm target}$ is the number of target nuclei in the detector fiducial volume,
 and $\phi$ is the integrated neutrino flux. 
 
The NOvA ND consists of 193 metric tons of a fully active tracking calorimeter constructed from polyvinyl chloride (PVC) cells filled with liquid scintillator.
The liquid scintillator is 62\% of the fiducial mass. 
The target nuclei for neutrino interactions are predominantly carbon (66.7\% by mass), chlorine (16.1\%) and hydrogen (10.8\%), with small contributions from titanium (3.2\%), oxygen (3.0\%) and other nuclei.
Each cell is 3.9\,cm wide, 6.6\,cm deep, and 3.9\,m in length.
Cells are arranged in planes alternating between horizontal and vertical orientations to provide three-dimensional reconstruction of neutrino interactions.
The fully active volume of the detector is 12.8\,m in length, consisting of 192 contiguous PVC planes with 96 cells each.  
Each plane is approximately $0.18$ radiation lengths. 	
Downstream of the fully active volume is a muon range stack with ten layers of 10-cm-thick steel plates interleaved with pairs of one vertical and one horizontal scintillator plane to enhance muon containment.
Scintillation light generated by charged particles passing through a cell is captured by a wavelength-shifting fiber connected to a Hamamatsu avalanche photodiode (APD) \cite{Hamamatsu:apd} at the end of the cell.
The APD signals are continuously digitized, and those above a preset threshold are recorded with associated time and charge. 

The NuMI neutrino beam is produced by colliding 120\,GeV protons from the main injector accelerator on a 1.2-m-long graphite target. 
Charged hadrons produced in the target are focused by two magnetic horns downstream of the target to select positive mesons which then decay into neutrinos in a 675\,m long decay pipe. 
This analysis uses data corresponding to 3.72$\times10^{20}$\,protons on target (POT). 
The neutrino beam is simulated by FLUKA \cite{Bohlen:2014fluka} and the FLUGG \cite{Campanella:1999flugg} interface to GEANT4 \cite{Agostinelli:2003geant4}. 
External thin-target hadron production measurements are used to correct and constrain the neutrino flux via the PPFX package developed for the NuMI beam by the MINERvA Collaboration \cite{Aliaga:2016flux}.

The NOvA ND is 1\,km from the neutrino source, 100\,m underground, and on average 14.6\,mrad away from the central axis of the neutrino beam.
The neutrino flux seen in the NOvA ND is a narrow-band beam peaked at 1.9\,GeV, with 68\% of the flux between 1.1 and 2.8\,GeV and a mean of 2.7\,GeV due to the high-energy tail.
The neutrino beam in the 0-120\,GeV energy region is predominantly $\nu_\mu$ (91\%), with a small contamination from $\nu_e$ (1\%) and antineutrinos (8\%).
In this measurement, the effect of antineutrinos in the flux is accounted for using simulation to give a solely neutrino-induced result.
The predicted integrated neutrino flux from 0 to 120\,GeV in the detector volume used in this analysis is $\phi_{\nu} = 123.2\pm11.6 \text{neutrinos}/\,\text{cm}^2/10^{10}$\,POT.

Neutrino interactions in the detector are simulated by the GENIE 2.10.4 neutrino event generator \cite{Andreopoulos:2010genie} except DFR. 
The Rein-Sehgal PCAC-based model is used to simulate the coherent process. 
To simulate NC RES and DIS events, the two major background contributions, the Rein-Sehgal model for baryon-resonance production \cite{Rein:1981resRS} and the Bodek-Yang model \cite{Bodek:2005by} are used. 
The only DFR model implemented in GENIE is the Rein model \cite{Rein:1986dfr}.  
However, the Rein model is valid only for the hadronic invariant mass $W>2$ GeV region, which is too high for the energy range of NOvA.  
To simulate the DFR background for this measurement, events are first generated by the Rein model in GENIE (v2.12.2) and then reweighted to an estimation based upon the PCAC calculation by Kopeliovich \textit{et al.} which includes both DFR and non-DFR contributions. \cite{Kopeliovich:2012}\cite{Kopeliovich:np}. 
In this estimation, Kopeliovich's prediction of $\frac{d\sigma}{d(|t|-|t|_{\text{min}})}$ for inclusive $\nu p \rightarrow \nu p \pi^0$ is fitted with GENIE (without DFR) and an exponential term, where $|t|_{\text{min}}$ is the minimum possible value of $|t|$.    
The exponential term extracted is considered to be the maximum possible contribution from DFR. 
The DFR events are simulated independently from RES and DIS, with the interference between them neglected, the effect of which is estimated and taken as systematic uncertainty. 
More details on the DFR modeling can be found in the Appendix. 
The cross section and selection efficiency of the DFR model used in this measurement are also provided in the Appendix, so that alternative models may be applied to estimate the impact on this measurement.

The nuclear model used in the simulation is the Bodek-Richie relativistic Fermi gas model with short-range nucleon-nucleon correlations \cite{Bodek:1981br1,Bodek:1981br2}. 
Final-state interactions of hadrons inside the nucleus are simulated in GENIE using an effective intranuclear cascade model \cite{Andreopoulos:2010genie}. 
GEANT4 \cite{Agostinelli:2003geant4} is used to simulate the detector's response to the final-state particles from neutrino interactions.
The propagation of photons produced by the simulated energy depositions, the response of the APDs, and the digitization of the resulting waveform is accomplished with a custom simulation package. 

In both data and simulation, the recorded cell signals (hits) in the NOvA detector are first collected into groups by their space and time information.  
Each collection of hits is assumed to come from a single neutrino interaction. 
The intersection of the particle paths found in the collection using a Hough transform \cite{Fernandes:2008hough} are taken as seeds to find the interaction vertex. 
Hits are further clustered into ``prongs'' with defined start points and directions emanating from the vertex.
Each prong contains hits attributed to one particle.

\begin{figure*}[ht]
\begin{center}
\includegraphics[width=0.49\linewidth]{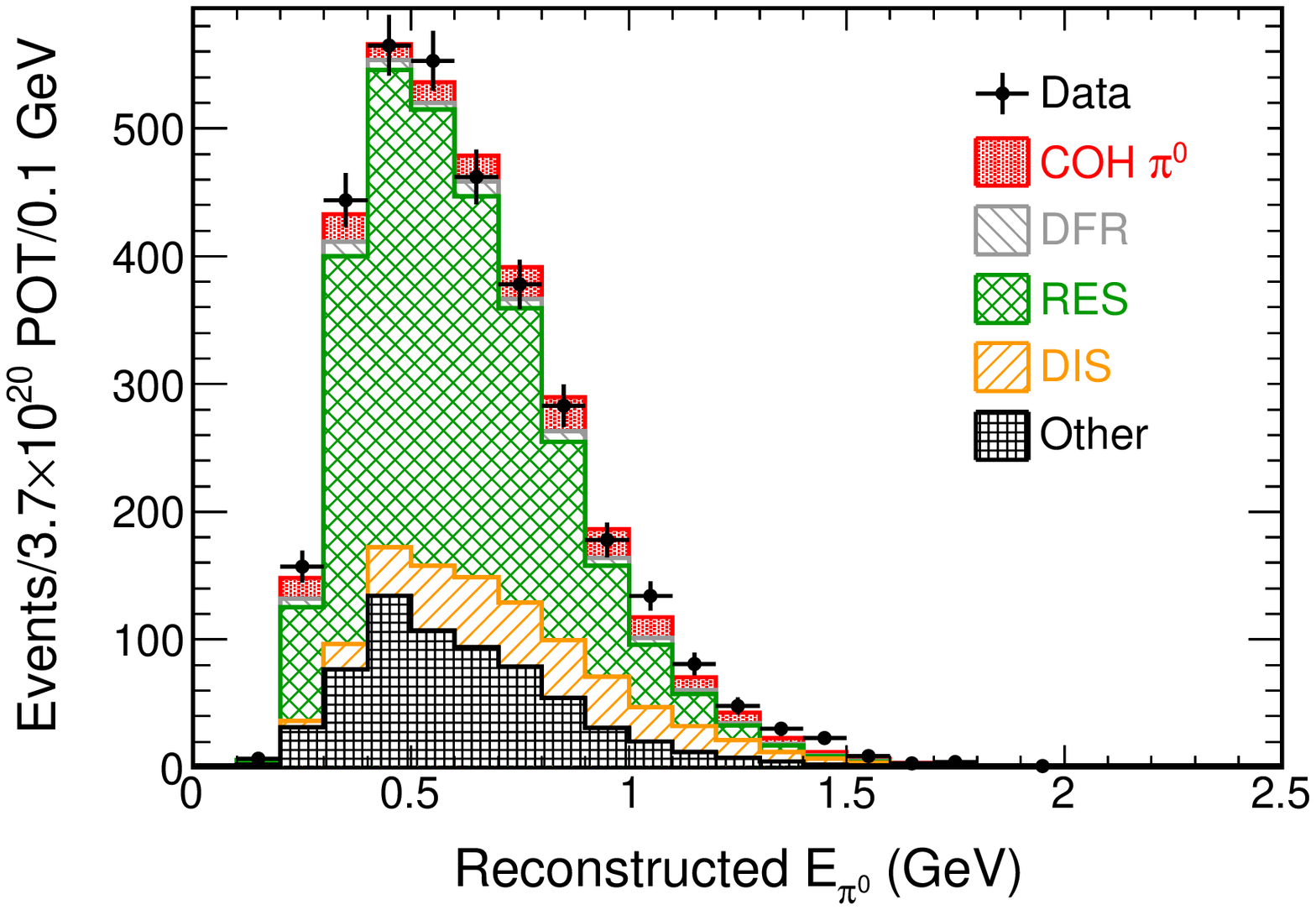}
\includegraphics[width=0.49\linewidth]{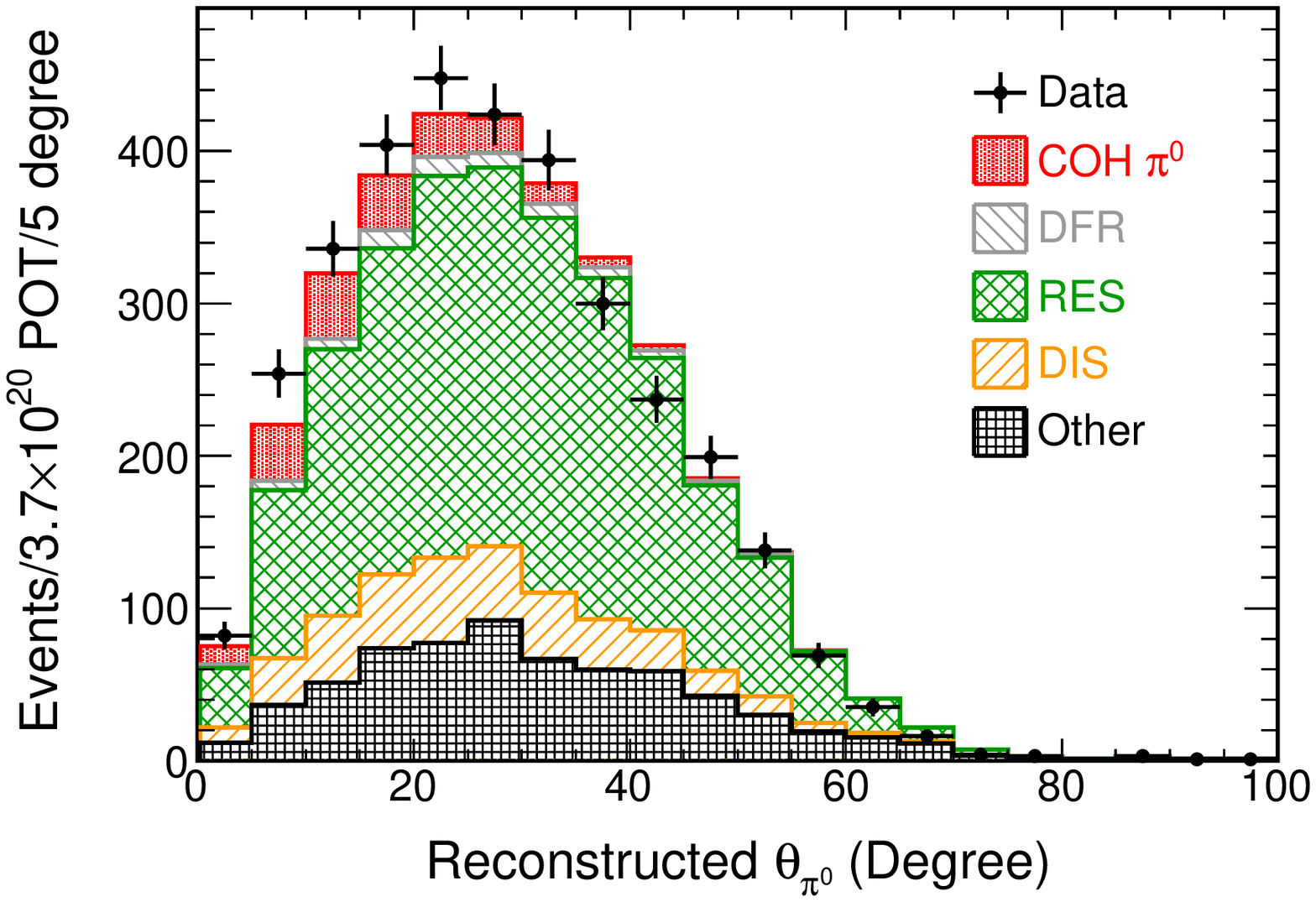}
\end{center}
\caption{Reconstructed $\pi^0$ energy (left) and angle with respect to beam (right) of the control sample events after the background fit.\\
\label{figure_egg_angle_ctr_norm}}
\end{figure*}

\begin{figure*}[ht]
\begin{center}
\includegraphics[width=0.49\linewidth]{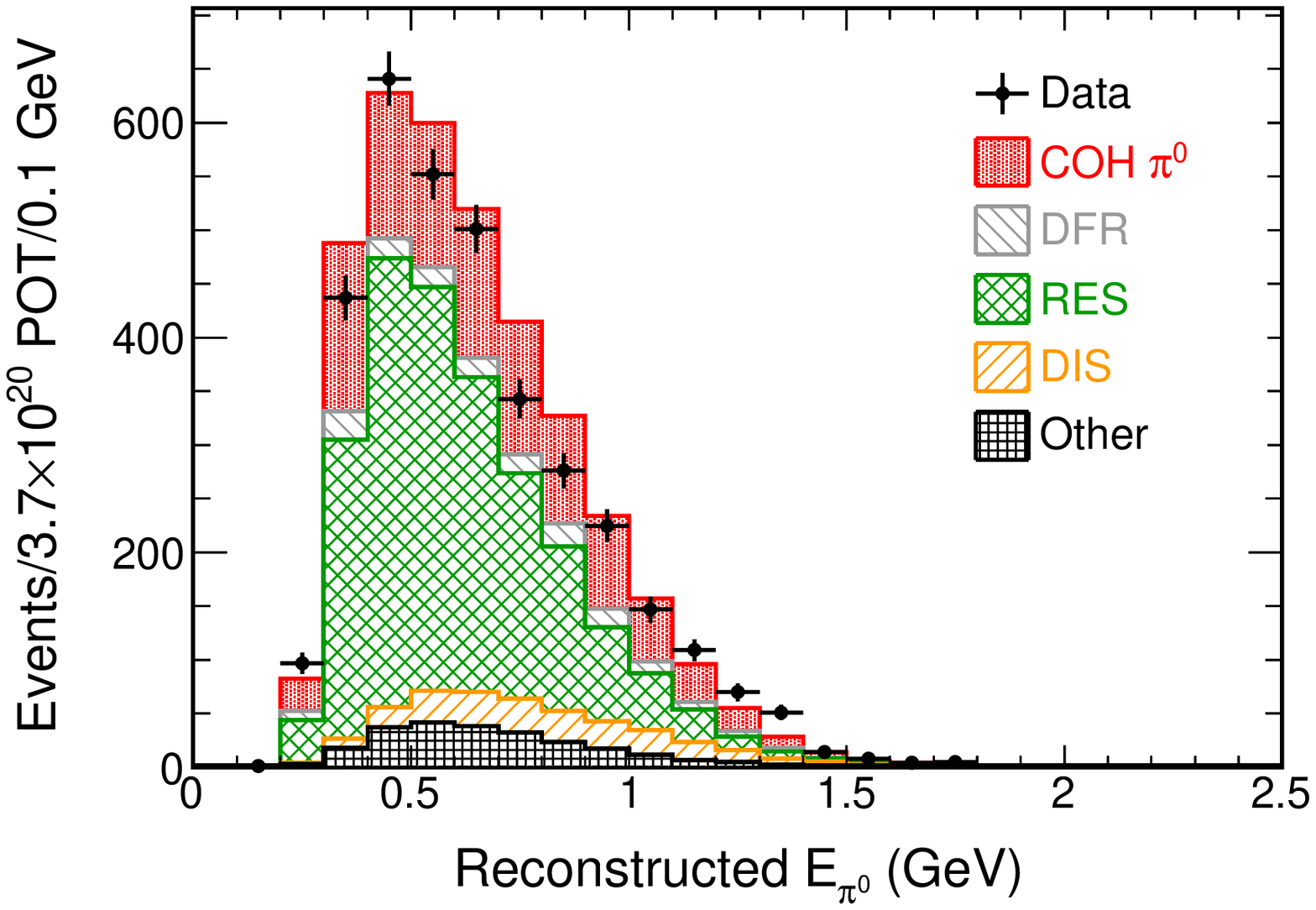}
\includegraphics[width=0.49\linewidth]{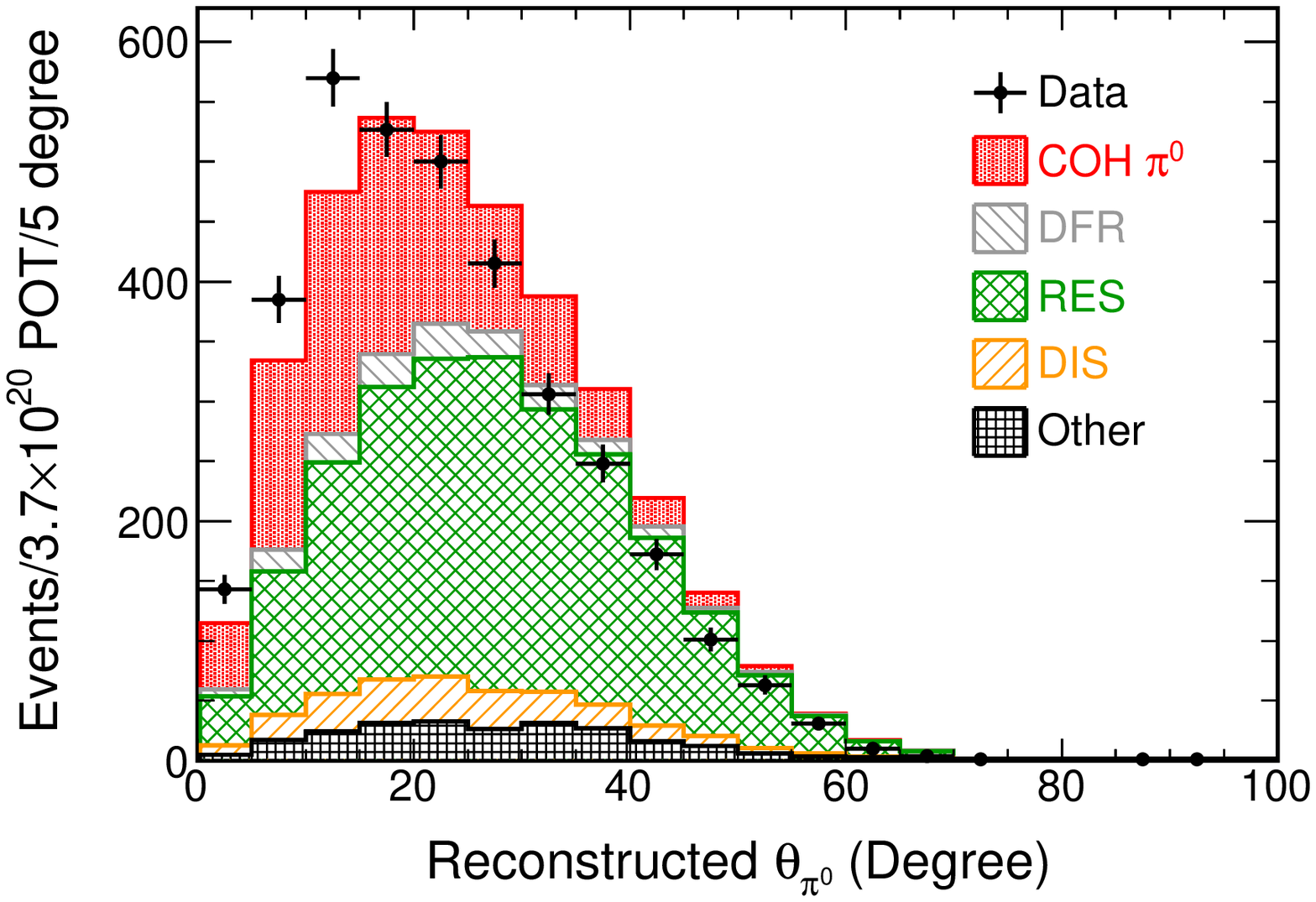}
\end{center}
\caption{Reconstructed $\pi^0$ energy (left) and angle with respect to beam (right) of the signal sample events. The simulated backgrounds are normalized by the control sample data.\\  
\label{figure_egg_angle_coh_norm}}
\end{figure*}

The events selected by this analysis are required to have exactly two reconstructed prongs contained in the fully active volume of the detector, 
both identified as electromagnetic-like showers    
by log-likelihood functions based upon $dE/dx$ information in both the longitudinal and transverse directions of the prongs \cite{Niner:2015thesis, Sachev:2016thesis}. 
A convolutional neural network trained for NC-CC separation \cite{Adamson:2017zcg} is used to reject CC events. 
The energy of the prong is calculated as the sum of the calibrated energy deposited in each cell. 
The invariant mass is calculated from the momenta and opening angle of the reconstructed prongs assuming both are photons, as 
\begin{equation}
M_{\gamma\gamma} = \sqrt{2E_{\gamma1}E_{\gamma2}(1-\cos\theta_{\gamma\gamma})} ,
\end{equation}
where $E_{\gamma1}$ and $E_{\gamma2}$ are the energies of the two prongs and $\theta_{\gamma\gamma}$ is the opening angle between them. 
The energy scales of data and simulated events are tuned independently so that the mass peaks match the $\pi^0$ mass (134.977\,MeV/c$^2$) \cite{Tanabashi:pdg2018}. 
Only events with reconstructed $\pi^0$ mass between 85 and 185\,MeV/c$^2$ are selected to reduce backgrounds.
The momenta of the two reconstructed prongs are summed up to obtain the reconstructed momentum of the $\pi^0$.  

As shown in Fig. \ref{figure_mggnc}, the selected events are high-purity NC $\pi^0$s (90\%), including both coherent signal and background arising from NC RES and DIS, with small contributions from DFR $\pi^0$ production and other interactions. 
The background events may have extra energy, especially in the vertex region, but not enough to be reconstructed as prongs.
To better control the background, the NC $\pi^0$ sample is further divided into two subsamples using kinematic variables: 
the ratio of the calorimetric energy included in the reconstructed $\pi^0$ to the total energy in the event ($E_{\pi^0}/E_{\rm Tot}$), 
and the energy in the vertex region defined as the first eight planes from the reconstructed interaction vertex ($E_{\rm Vtx}$).
The signal-enhanced sample is defined as events with most of their energy in the two photon prongs ($E_{\pi^0}/E_{\rm Tot}>0.9$) and low vertex energy ($E_{\rm Vtx}<0.3$\,GeV) to include most of the coherent signal and reduce background.
The rest of the events are defined as a control sample, dominated by $\pi^0$s produced by RES and DIS interactions.
The signal and control sample selection is shown in Fig. \ref{figure_caleratio_vtxe}. 

\begin{figure*}[ht]
\begin{center}
\includegraphics[width=0.49\linewidth]{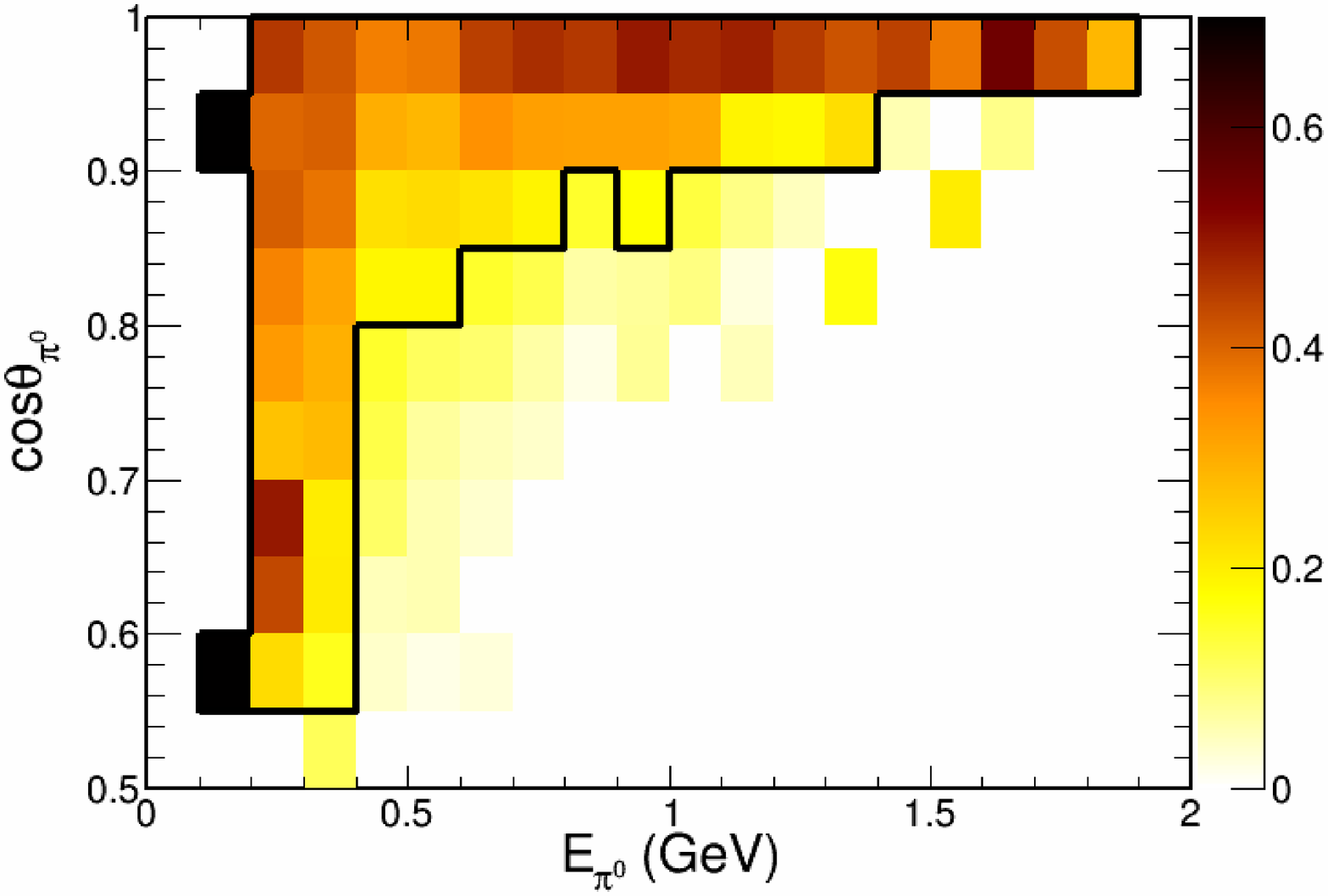}
\includegraphics[width=0.49\linewidth]{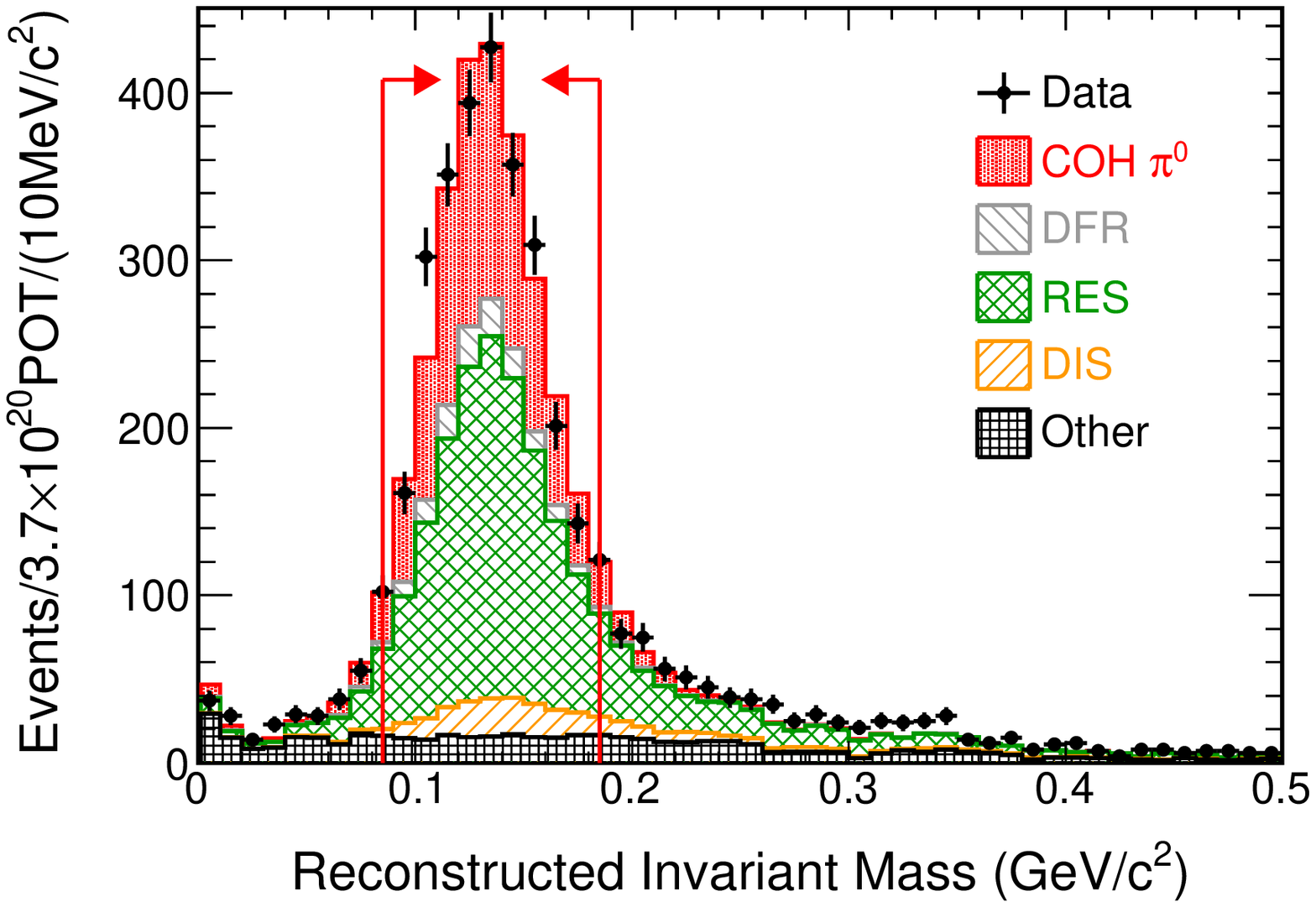} 
\end{center}
\caption{Left: Ratio of coherent $\pi^0$ signal to total simulated events in the signal sample in the 2D space of $\pi^0$ energy and $\cos\theta$. The region inside the lines is the coherent region defined as bins with $>15\%$ of total simulated events being coherent $\pi^0$. 
Right:  $\pi^0$ invariant mass of the signal sample events in the coherent region as described by the left plot with the background normalized by the control sample data.
Vertical lines with arrows show the range of invariant masses accepted into the analysis.
\label{figure_cohr}}
\end{figure*}

The control sample data are used to constrain the background prediction. 
The simulated distributions of RES and DIS events in the $\pi^0$ energy and angle ($\cos\theta$ with respect to the average beam direction) 2D space are used as templates and scaled to fit the control sample data. 
RES and DIS have distinct $\pi^0$ energy and angle distributions, and together they account for approximately $90\%$ of the total background.
The fitting parameters are the normalization factors of the templates.
The other background components are kept fixed in the fit.
The fit results in an increase of the selected RES background by 17.5$\pm$6.2\% and a decrease in the DIS background by 43.1$\pm$13.8\%. 
The two fitting parameters are strongly anticorrelated.
The fit result is applied as a renormalization to the background in the signal sample.
It also provides a constraint on the systematic sources affecting backgrounds, which will be discussed later.
The energy and angle of the $\pi^0$s in the control sample and the signal sample with the renormalized backgrounds are shown in Figs. \ref{figure_egg_angle_ctr_norm} and \ref{figure_egg_angle_coh_norm}.
There are notable discrepancies between the signal sample data and simulation, especially in the $\pi^0$ angular distribution (Fig. \ref{figure_egg_angle_coh_norm}, right).  
The $\theta_{\pi^0}$ spectrum in the data favors production at angles closer to the beam direction than does the simulation, 
suggesting that the extrapolation from the $Q^2=0$ PCAC approximation to nonzero $Q^2$ values in the Rein-Sehgal model needs refinement. 
Similar discrepancies in pion angular distributions have been reported by the MINERvA experiment in recent measurements of charged-current coherent pion production \cite{Higuera:2014minerva,Mislivec:2017qfz}.
Further study of systematic uncertainties is ongoing to quantitatively address the discrepancies. 

\begin{figure*}[ht]
\begin{center}
\includegraphics[width=0.5\textwidth]{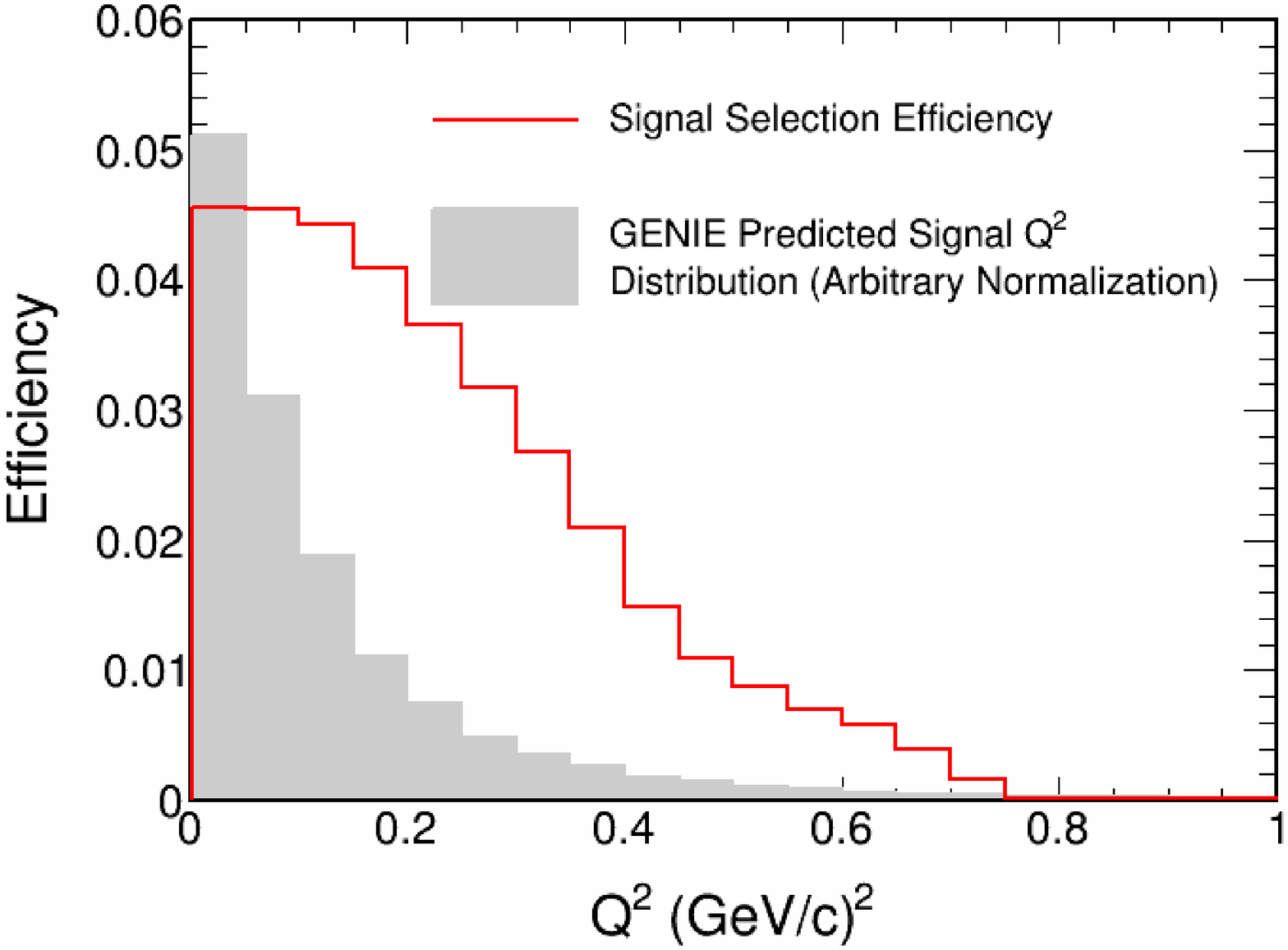}
\caption{Selection efficiency of NC coherent $\pi^0$ signal as a function of $Q^2$. The GENIE predicted signal $Q^2$ shape is shown in gray with arbitrary normalization. 
\label{figure_effq2}}
\end{center}
\end{figure*}

\begin{table*}
\caption{List of systematic and statistical uncertainties. \label{table_uncertainty}}
\begin{center}
\begin{tabular}{lc}
\hline 
Source & Measurement uncertainty (\%)\\ 
\hline 
Calorimetric energy scale & 3.4\\ 
Background modeling & 12.3\\ 
Coherent modeling & 3.7\\
Photon shower response & 1.1\\ 
External events & 2.4\\
Detector simulation & 2.0\\ 
Flux  & 9.4\\  
\hline 
Total systematic uncertainty & 16.6\\ 
Statistical uncertainty & 6.8 \\
\hline 
Total uncertainty& 17.9\\
\hline 
\end{tabular} 
\end{center}
\end{table*}

\begin{figure*}[ht]
\begin{center}
\includegraphics[width=0.49\textwidth]{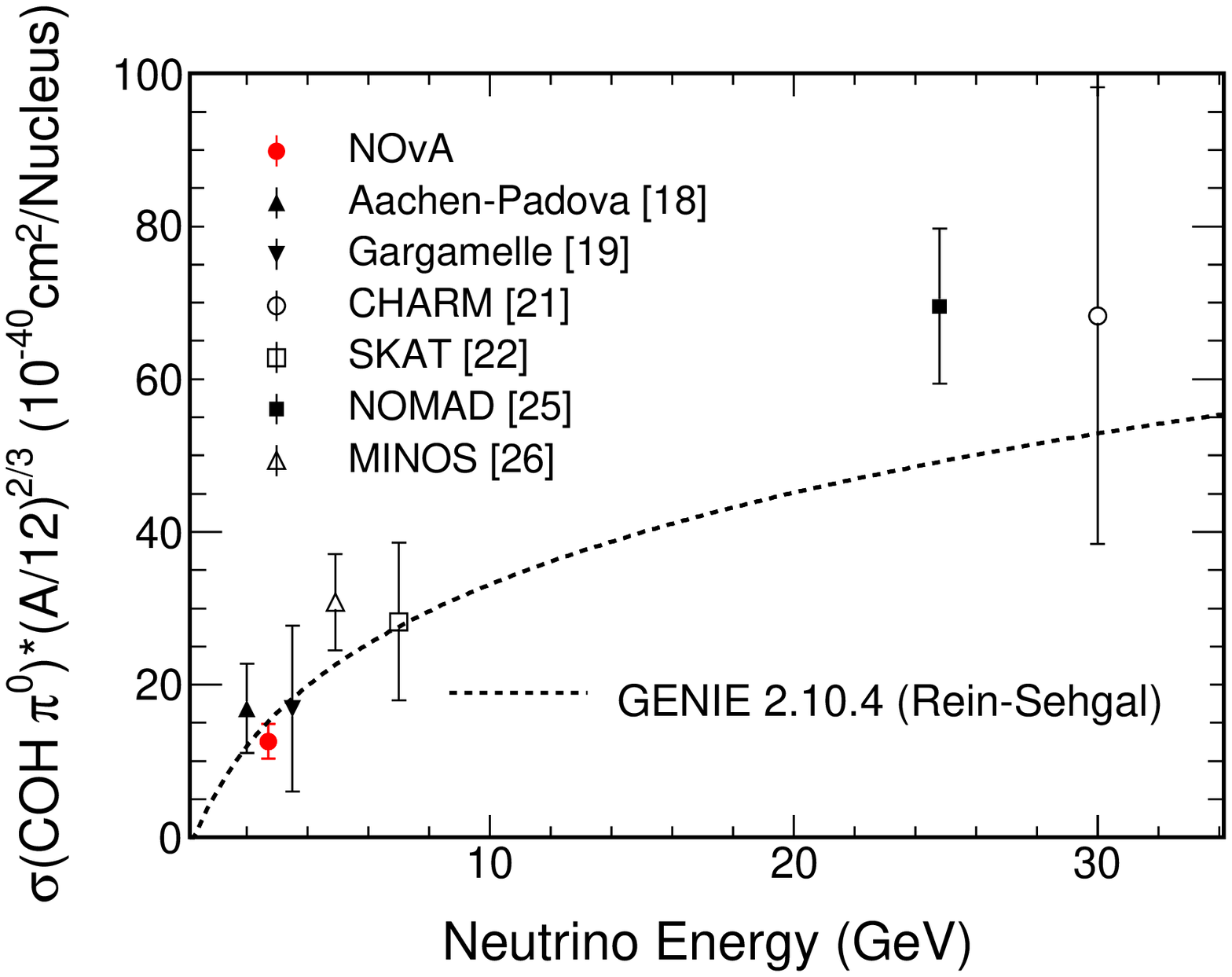}
\includegraphics[width=0.49\textwidth]{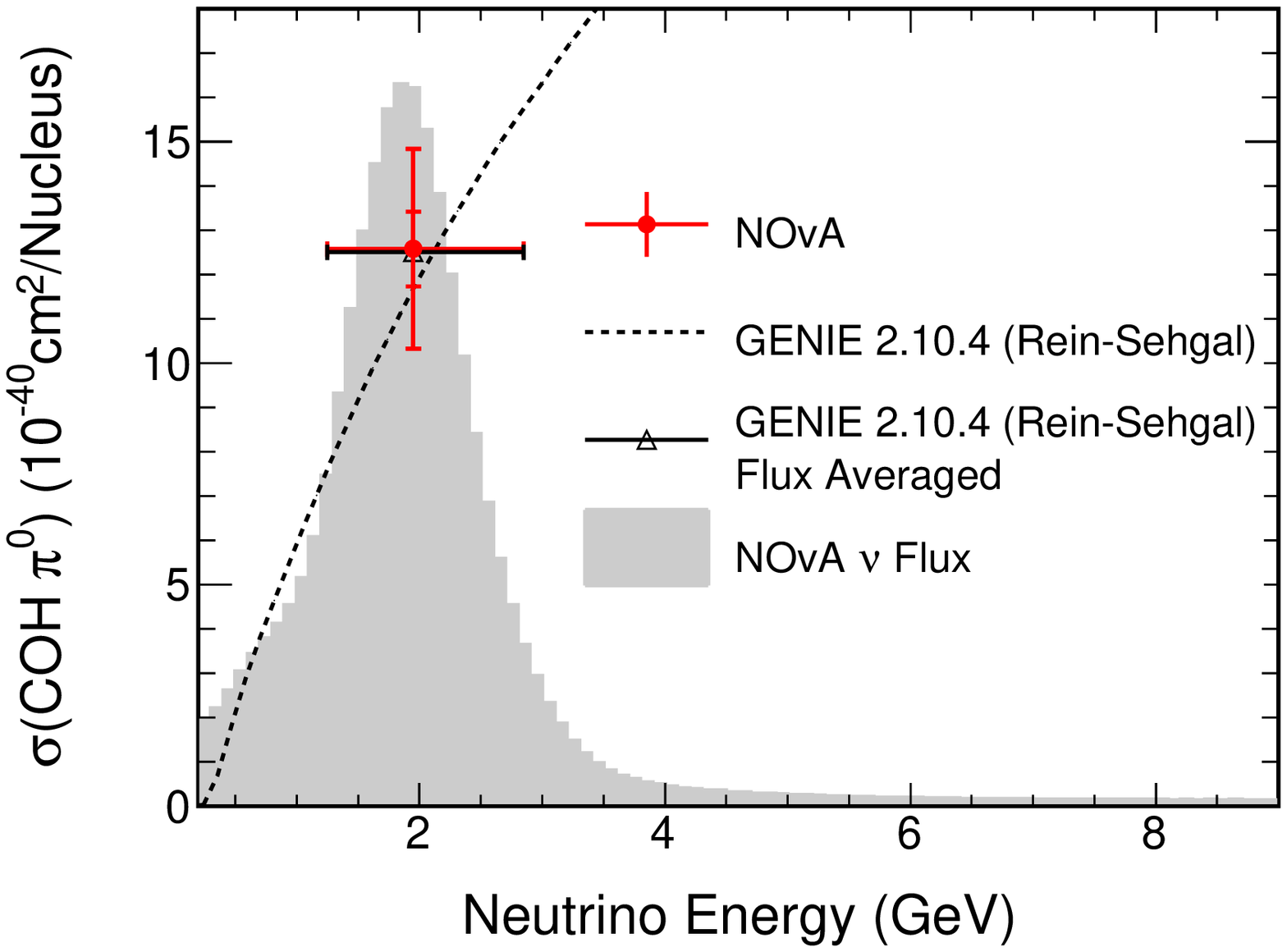}
\caption{Flux-averaged cross section of the NOvA NC coherent $\pi^0$ measurement. The left plot compares this measurement to previous measurements.  
The neutrino energy values of the NOvA data point and other measurements are represented by an average neutrino energy.
All results are scaled to a carbon target by a factor of $(A/12)^{2/3}$ following the Berger-Sehgal model approximation, where $A$ is the effective atomic number of the experiment. 
The dashed curve shows the GENIE prediction for a carbon target. 
The right plot compares this measurement with the GENIE predicted flux-averaged cross section from the Rein-Sehgal model. In this plot the neutrino energy of the NOvA data point is the median neutrino energy, and the horizontal error bar contains 68\% of the total neutrinos. The statistical uncertainty and statistical plus systematic uncertainty are shown as vertical error bars for the NOvA result. The GENIE prediction is shown both as a function of neutrino energy, and as a flux-averaged cross section.  The NOvA flux is shown in gray with arbitrary normalization. 
\label{figure_xsec}}
\end{center}
\end{figure*}

A coherent region in the 2D $\pi^0$ energy and angle space is defined as those bins with $>15\%$ predicted coherent $\pi^0$ signal purity (Fig. \ref{figure_cohr}, left). 
The selection is intentionally set loosely to reduce potential systematic uncertainties caused by the discrepancies in the $\pi^0$ kinematic distributions mentioned previously.
The invariant mass of the signal sample events is shown in Fig. \ref{figure_cohr}, right.
The signal selection efficiency is $4.1\%$ according to simulation.
Figure \ref{figure_effq2} shows the selection efficiency as a function of $Q^2$ along with the GENIE predicted signal $Q^2$ shape. 
Alternative coherent models may be applied to estimate the impact on this measurement with the selection efficiency provided.

The normalized background in this coherent region is subtracted from data to obtain the number of measured signal events. 
The number of simulated signal events is then normalized to the number extracted from the data.  
The calculation is iterated until the resulting changes in the estimated signal and background populations become negligible.
The outcome of this procedure is the coherent signal content, estimated to be $977\pm67$\,(stat) events. 
Neutrino- and antineutrino-induced coherent $\pi^0$s are indistinguishable in this measurement. GENIE predicts 94\% of the signal being neutrino induced. This percentage is used to correct the measurement to solely neutrino-induced.

The systematic uncertainties for this analysis arise from the calorimetric energy scale,  background modeling,  coherent signal modeling, 
detector response to photon showers,
detector simulation, particles entering the detector from external sources, and the simulation of neutrino flux. 
Data-driven methods are used wherever possible to establish the uncertainties. 

The calorimetric energy scale is constrained to within 1\% by the $\pi^0$ invariant mass distributions of simulation and data
which corresponds to a 3.4\% uncertainty on the cross-section measurement. 
The background-related uncertainty is constrained by the control sample data through the template fit method. 
The variations that can arise with the template fit to background are estimated by varying the background-modeling parameters within their $\pm1\sigma$ ranges as assigned by GENIE and then repeating the template fit.
The uncertainty from each background-modeling parameter is defined as the maximum change in the measured signal events.
To estimate the uncertainty from DFR modeling, the template fit is repeated with DFR added as an additional template with its normalization allowed to float. 
The fit gives DFR normalization factor $0.0+0.8$, which favors no DFR contribution. 
$-100\%$ uncertainty on DFR is assigned based upon this study.   
Another potential uncertainty source from DFR is the interference between DFR and other pion production channels on hydrogen. 
To estimate this effect, an additional systematic variation is created by reweighting all the $\nu p \rightarrow \nu p \pi^0$ events on hydrogen to the Kopeliovich model prediction, including both DFR and non-DFR.
The vertex energy cut used to define the signal sample and control sample is subject to nuclear effects which are not well modeled by GENIE.
To check the impact, the control sample is redefined by applying the same vertex energy ($E_{\rm Vtx}<0.3$\,GeV) as the signal sample so that the effect of potential mismodeling of vertex energy cancels out. 
The resulting difference in the measurement from the nominal is added to the systematic uncertainty from GENIE background modeling.

The uncertainty in the coherent signal modeling results in an uncertainty of the efficiency correction. 
This effect is evaluated by varying the modeling parameters in the Rein-Sehgal model: axial mass ($M_{A}$, $\pm50\%$) and nuclear radius ($R_{0}$, $\pm20\%$) \cite{Rein:1983rs,Rein:2007rs,Andreopoulos:2010genie}. 
To check the effect of the discrepancies in $\pi^0$ kinematic distributions on the total cross-section measurement, 
a test is performed by reweighting the simulated signal to data 
and comparing to the result obtained before reweighting.
A 1\% difference is found, which is negligible compared to the signal modeling uncertainty assigned.
Bremsstrahlung showers induced by energetic muons from external sources provide a data-driven constraint on the simulation of detector response to photon showers. 
Those bremsstrahlung showers are identified and the muons are removed to create a single photon control sample in the data and simulation \cite{Duyang:2015mrbrem}. 
The sample is subject to the same selection cuts as the $\pi^0$ photons, and the uncertainty is evaluated as the 1\% difference between the data and simulation in selection efficiency.
Lastly, the neutrino flux uncertainty comes from beam focusing and hadron production with external thin-target hadron production data constraints applied \cite{Aliaga:2016flux}. 
The systematic sources and uncertainties are summarized in Table \ref{table_uncertainty}.
The dominant sources are background modeling and flux uncertainties.
The total systematic uncertainty is estimated to be 16.6\%. 

\begin{table*}
\begin{center}
\caption{Summary of NC coherent $\pi^0$ measurements. The effective atomic number ($A$) and the average neutrino energy ($\langle E_{\nu} \rangle$) are shown for each experiment. The results are reported as total cross section per nucleus, cross-section ratios to inclusive $\nu_{\mu}$-CC or to the prediction of Rein-Sehgal model. 
\label{table_worlddata} }
\footnotesize
\begin{tabular}{cccccc}
\hline 
Experiments & $A$ \footnote{The effective atomic number calculations may differ between experiments.}& $\langle E_{\nu} \rangle$ (GeV) & $\sigma$ ($10^{-40}\,\text{cm}^2/N$) & $\sigma$/$\sigma$($\nu_{\mu}$-CC) & $\sigma$/$\sigma$(Rein-Sehgal)  \footnote{The implementaions of the Rein-Sehgal model used by other experiments (MiniBooNE\cite{Aguilar-Arevalo:2008miniboone} and SciBooNE\cite{Kurimoto:2008sciboone}) could be considerably different from the GENIE implementation used by the NOvA measurement. A comparison of the Rein-Sehgal predictions of CC coherent in different generators can be found in Ref. \cite{Higuera:2014minerva}.}
\\ 
\hline 
Aachen-Padova \cite{Faissner:1983ap} & 27 & 2 & 29$\pm$10 & • & •\\ 
Gargamelle \cite{Isiksal:1984Gargamelle} & 31 & 3.5 & 31$\pm$20 & • & •\\
CHARM \cite{Bergsma:1985charm} & 20 & 30 & 96$\pm$42 & • & •\\  
SKAT \cite{Grabosch:1986SKAT}& 30 & 7 & 79$\pm$28 &4.3$\pm$1.5& •\\  
15' BC \cite{Baltay:1986bc} & 20 & 20 & • & 0.20$\pm$0.04 & \\  
NOMAD \cite{Kullenberg:2009nomad} & 12.8 & 24.8 & 72.6$\pm$10.6 & 3.21$\pm$0.46 & •\\ 
MiniBooNE \cite{Aguilar-Arevalo:2008miniboone} & 12 & 0.8 & • & • & 0.65$\pm$0.14\\ 
SciBooNE \cite{Kurimoto:2008sciboone} & 12 & 0.8 & • & • & 0.9$\pm$0.20\\   
MINOS \cite{Adamson:2016minos} & 48 & 4.9& 77.6$\pm$15.9 & • & • \\  
NOvA & 13.8 & 2.7 & 13.8$\pm$2.5 & • & •\\   
\hline 
\end{tabular} 
\end{center}
\end{table*}


The flux-averaged cross section of NC coherent $\pi^0$ production in this measurement is calculated using Eq. (2).
The measured cross section is $\sigma = 13.8\pm0.9 (\text{stat})\pm2.3 (\text{syst}) \times 10^{-40}\,\text{cm}^2/\text{nucleus}$ at the average neutrino energy of 2.7\,GeV. 
The effective atomic number $A = 13.8$ is calculated as 
\begin{equation}
A = \left( \sum_{i}{\frac{n_i}{n_{Tot}}A_i^{2/3}}\right)^{3/2} ,
\end{equation}
where $A_i$ is the atomic number of each chemical element (excluding hydrogen) and $\frac{n_i}{n_{Tot}}$ is its fraction to the total number of nuclei in the fiducial volume. 
The factor $A_i^{2/3}$ is an approximate cross-section scaling between different nuclei in accordance with the Berger-Sehgal model \cite{Berger:2009bs}. 
Other models may differ in the prediction of $A$ dependence of coherent pion production. 

Figure \ref{figure_xsec} and Table \ref{table_worlddata} show this measurement together with other measurements and the GENIE prediction. 
All measurements in Fig. \ref{figure_xsec} are scaled to a carbon target by a scale factor of $(A/12)^{2/3}$ for the purpose of comparison. 
The flux-averaged NC coherent $\pi^0$ cross section of this work is in agreement with the cross-section prediction of the Rein-Sehgal model (GENIE implementation), 
although some discrepancies in the $\pi^0$ kinematic distributions are observed.
This result is the most precise measurement of NC coherent $\pi^0$ production in the few-GeV neutrino energy region, 
and the first such measurement on a carbon-dominated target in this energy range. 
It benefits both current and future long-baseline neutrino oscillation experiments in background prediction with reduced uncertainty.

\section*{Acknowledgments}

This document was prepared by the NOvA Collaboration using the resources of the Fermi National Accelerator Laboratory (Fermilab), a U.S. Department of Energy, Office of Science, HEP user facility. Fermilab is managed by Fermi Research Alliance, LLC (FRA), acting under Contract No. DE-AC02-07CH11359. 
This work was supported by the U.S. Department of Energy; the U.S. National Science Foundation; the Department of Science and Technology, India; the European Research Council; the MSMT CR, GA UK, Czech Republic; the RAS, RFBR, RMES, RSF, and BASIS Foundation, Russia; CNPq and FAPEG, Brazil; STFC and the Royal Society, United Kingdom; and the state and University of Minnesota.  We are grateful for the contributions of the staffs of the University of Minnesota at the Ash River Laboratory and of Fermilab.
\vspace*{-5mm}

\section*{APPENDIX: DIFFRACTIVE PION PRODUCTION}

\begin{figure*}[ht]
\begin{center}
\includegraphics[width=0.49\textwidth]{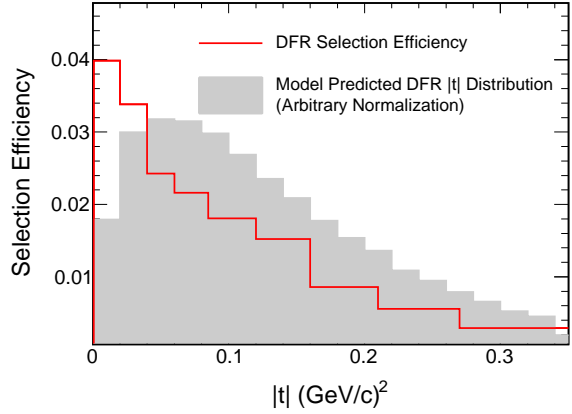}
\caption{Selection efficiency of DFR as a function of $|t|$. The DFR $|t|$ distribution estimated from Kopeliovich \textit{et al.} is shown in gray with arbitrary normalization.
\label{figure_dfr_eff}}
\end{center}
\end{figure*}

NC DFR pion production on free protons (hydrogen) is a background process to the coherent signal. 
It produces a forward-going pion with small momentum transfer to the recoil proton and becomes indistinguishable from coherent if the recoil proton is undetected.  
The recoil protons, when detected, could create additional prongs, increase the vertex energy, or decrease the ratio of $E_{\pi^0}/E_{Tot}$, causing the DFR events failing the selection cuts as a result.  
The acceptance of DFR, therefore, depends upon the kinetic energy of the recoil proton ($T_p$), which is related to $|t|$ by $T_p = |t|/2m_p$. 
The DFR selection efficiency in this measurement is shown in Fig. \ref{figure_dfr_eff} as a function of $|t|$. 
It is notable that the selection efficiency decreases with $|t|$, since the proton energy increases with $|t|$, and the overall efficiency (1.7\%) is considerably lower than the coherent signal (4.1\%).

\begin{figure*}[ht]
\begin{center}
\includegraphics[width=0.49\textwidth]{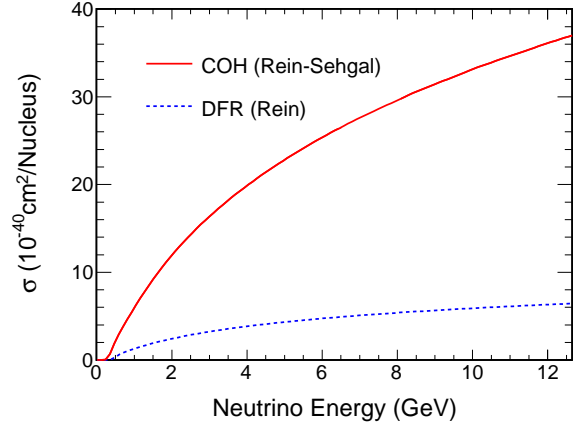}
\caption{
Cross section of DFR $\pi^0$ production on hydrogen as a function of incoming neutrino energy predicted by the Rein model (GENIE 2.12.2),  compared with coherent $\pi^0$ production on carbon predicted by the Rein-Sehgal model (GENIE 2.10.4). 
\label{figure_dfr_rein}}
\end{center}
\end{figure*}

\begin{figure*}[ht]
\begin{center}
\includegraphics[width=0.49\textwidth]{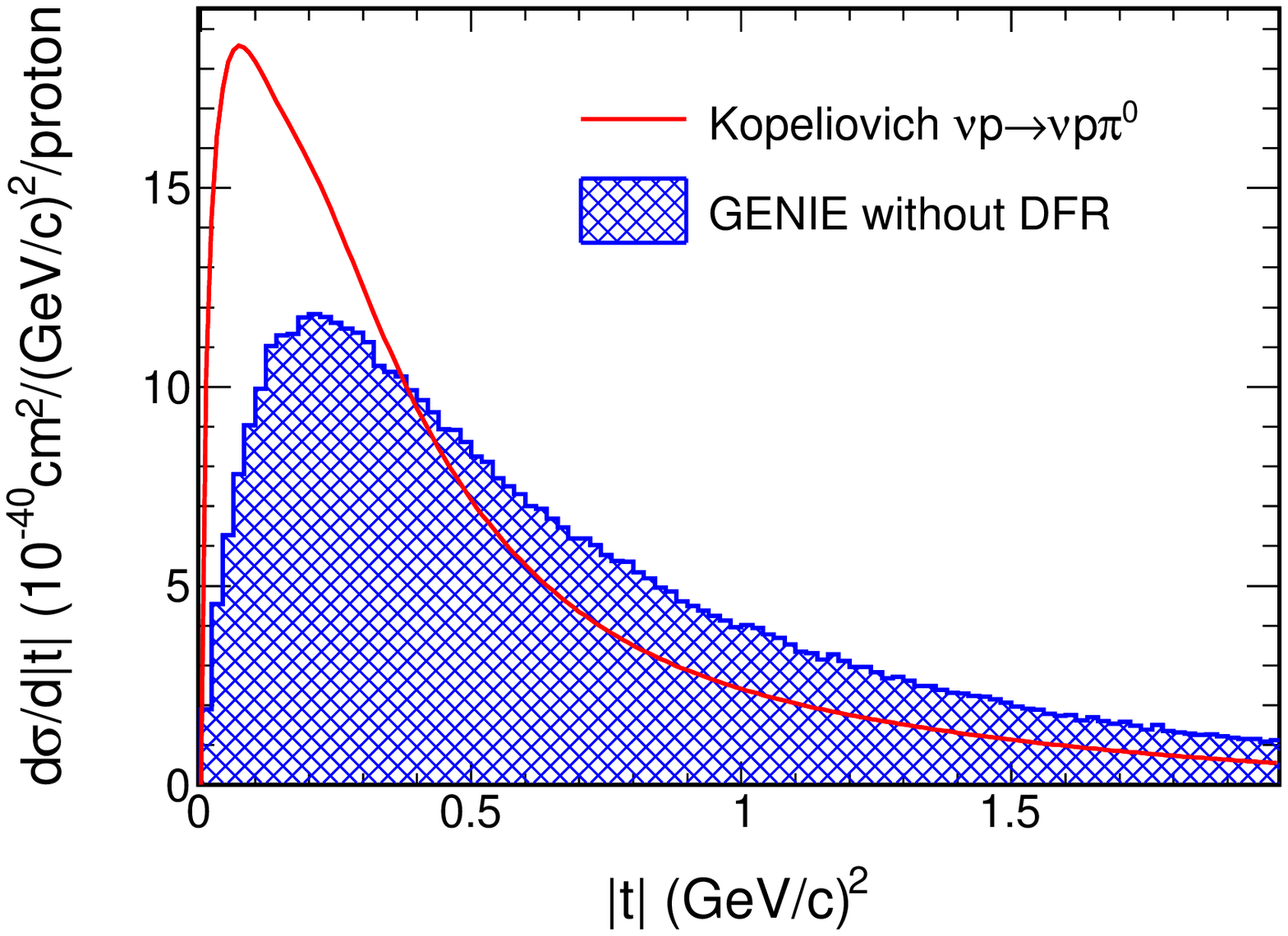}
\includegraphics[width=0.49\textwidth]{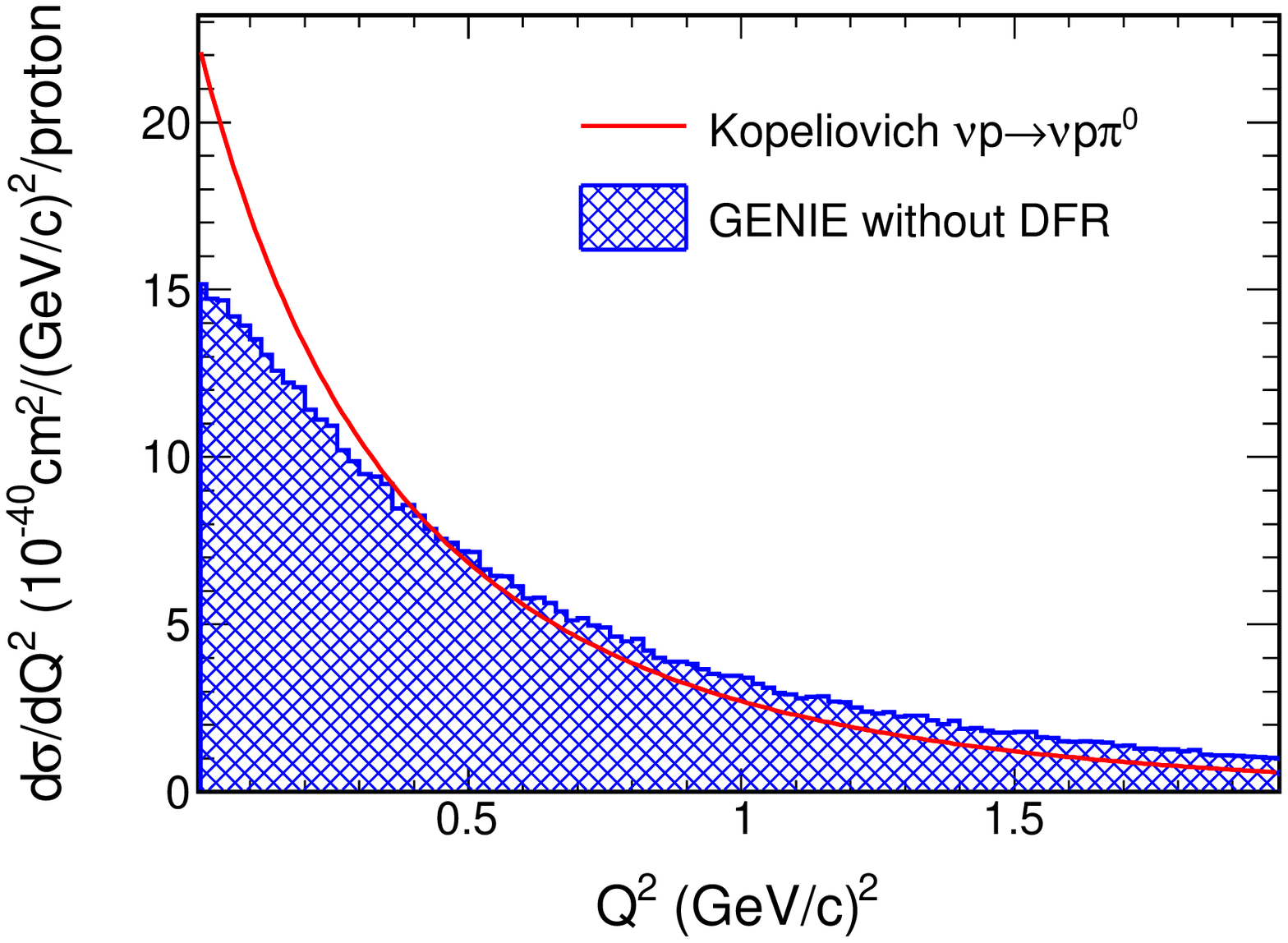}
\caption{The Kopeliovich $\frac{d\sigma}{d|t|}$ (left) and $\frac{d\sigma}{dQ^2}$ (right) predictions of $\nu p \rightarrow \nu p \pi^0$ at $E_\nu = 2.7$\,GeV, comparing with GENIE 2.12.2 prediction of the same channel without DFR at the same energy (shape only).  Enhancements in the low-$Q^2$ and low-$t$ region can be observed from Kopeliovich. \label{figure_dfr_q2_t}}
\end{center}
\end{figure*}

\begin{figure*}[ht]
\begin{center}
\includegraphics[width=0.49\textwidth]{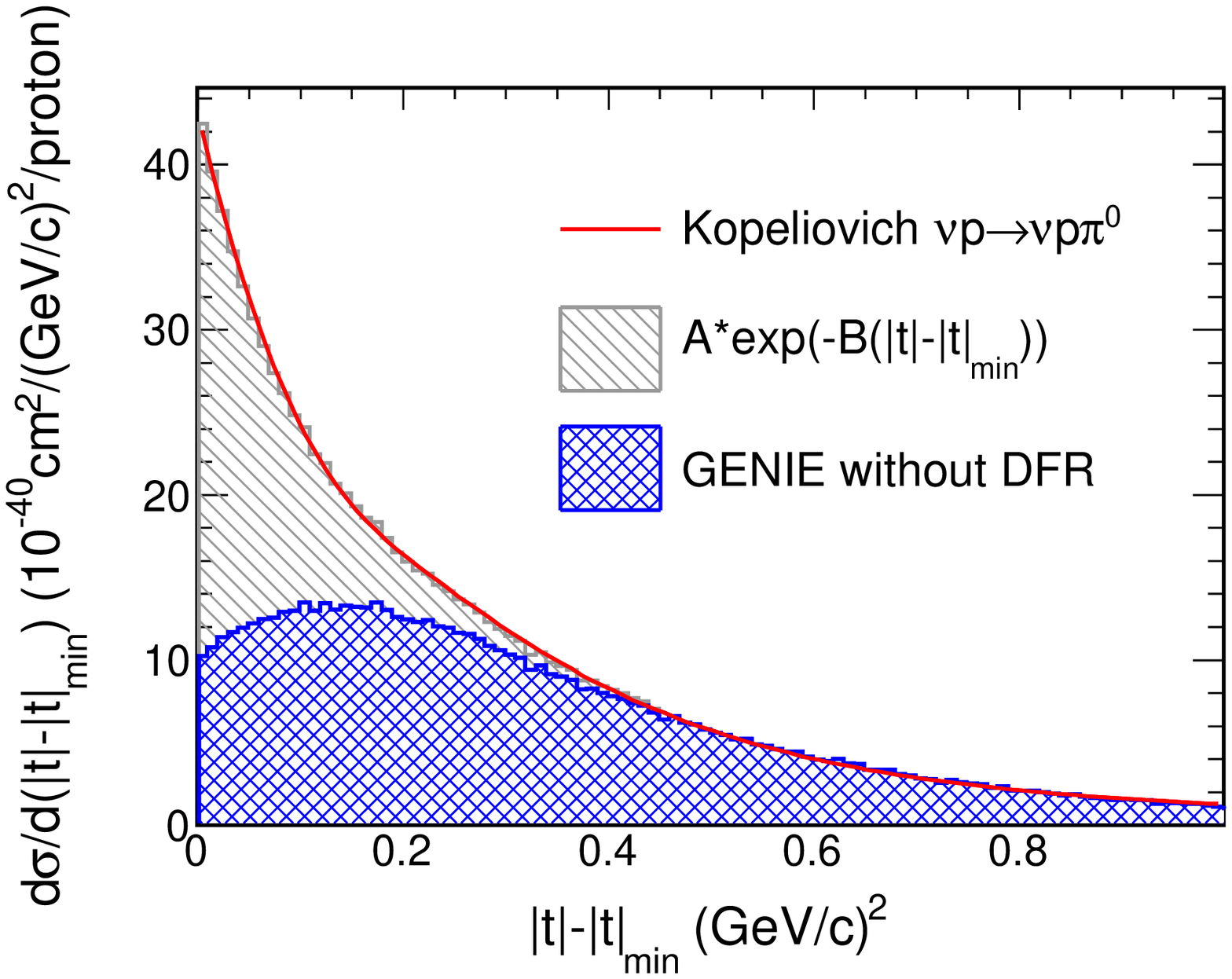}
\includegraphics[width=0.49\textwidth]{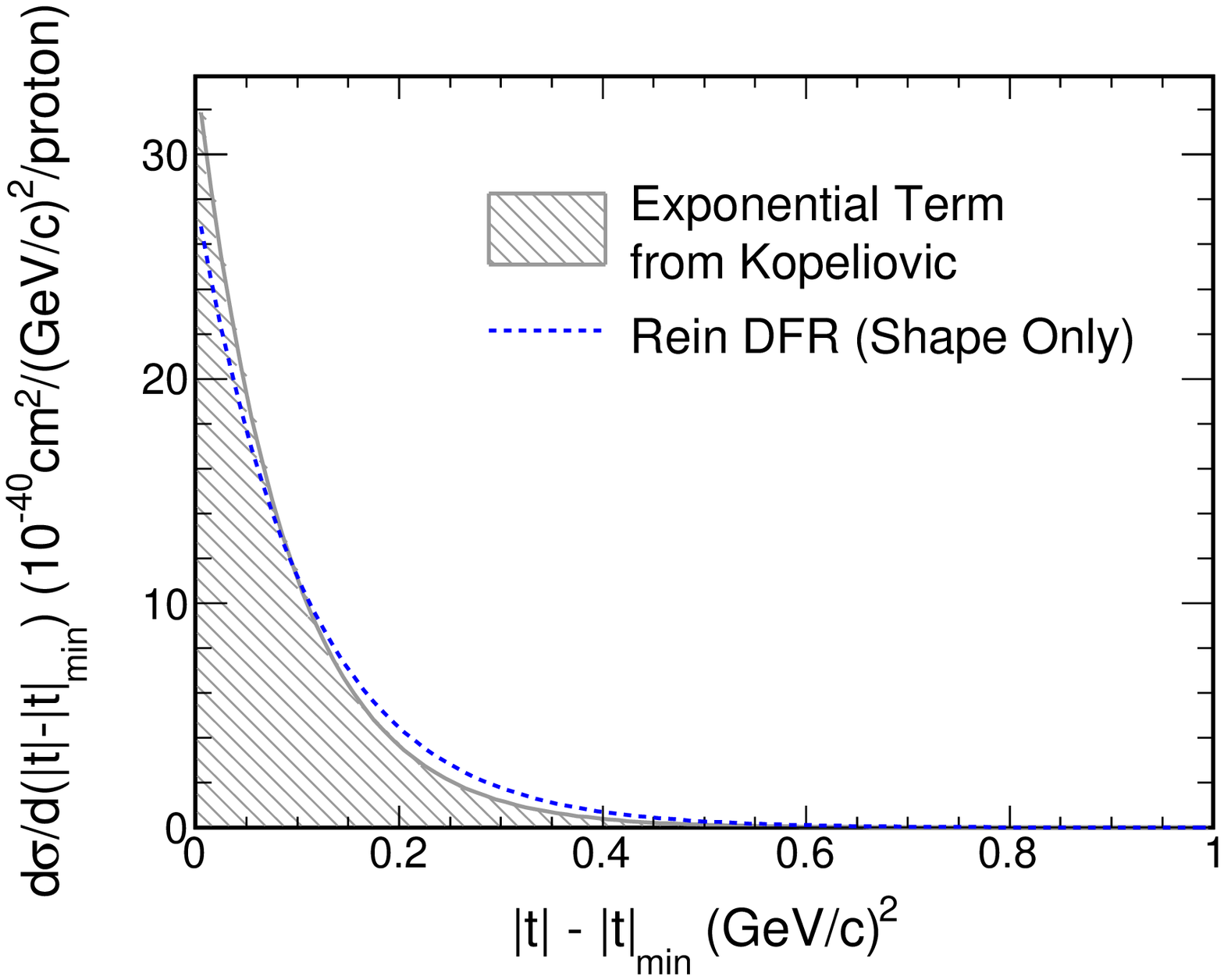}
\caption{Left: The Kopeliovich $\frac{d\sigma}{d(|t|-|t|_{\text{min}})}$ prediction of $\nu p \rightarrow \nu p \pi^0$ at $E_\nu = 2.7$\,GeV fitted with the GENIE 2.12.2 prediction of the same channel without DFR and an exponential term. The fitted exponential term is considered to be the maximum estimation of DFR $\frac{d\sigma}{d(|t|-|t|_{\text{min}})}$. Right: A shape comparison between the fitted exponential term from Kopeliovich and the term extracted from the Rein model prediction for DFR. A slightly softer shape is observed for the one from Kopeliovich. 
\label{figure_dfr_tmtmin}}
\end{center}
\end{figure*}

DFR is simulated by the Rein model in GENIE 2.12.2 and is predicted to be about $20\%$ of the coherent cross section on carbon in the few-GeV energy region (Fig. \ref{figure_dfr_rein}).  
However, the Rein model is intended for the hadronic invariant mass $W>2$\,GeV region.
In the $W<2$\,GeV region, the interference between DFR and RES or non-RES pion productions makes the performance of the Rein model questionable. 
Alternatively, the DFR cross section can be estimated by a similar method as in Ref. \cite{Mislivec:2017qfz} from the calculation of inclusive $\nu p \rightarrow \nu p \pi^0$ by Kopeliovich \textit{et al.}, which is based upon the PCAC hypothesis and includes both DFR and non-DFR contributions \cite{Kopeliovich:2012}\cite{Kopeliovich:np}.   
In Fig. \ref{figure_dfr_q2_t}, the left shows Kopeliovich's predictions of $\nu p \rightarrow \nu p \pi^0$ in $\frac{d\sigma}{d|t|}$ at NOvA's average neutrino energy (2.7\,GeV). 
This prediction is compared with the GENIE 2.12.2 prediction of $\nu p \rightarrow \nu p \pi^0$ without DFR, and an enhancement is observed in the low-$|t|$ region.  
A similar enhancement can be observed at low $Q^2$ (Fig. \ref{figure_dfr_q2_t}, right). 
The DFR contribution to $\nu p \rightarrow \nu p \pi^0$ can be quantified by fitting Kopeliovich's prediction of $\frac{d\sigma}{d(|t|-|t|_{\text{min}})}$ 
with GENIE without DFR 
plus an exponential term $A*\exp(-B(|t|-|t|_{\text{min}}))$, where $A$ and $B$ are fitting parameters (Fig. \ref{figure_dfr_tmtmin}, left). 
$\frac{d\sigma}{d(|t|-|t|_{\text{min}})}$ is used instead of $\frac{d\sigma}{d|t|}$ since the DFR $\frac{d\sigma}{d(|t|-|t|_{\text{min}})}$ follows an exponential form, while $\frac{d\sigma}{d|t|}$ deviates from an exponential at low $|t|$ because of the $|t|_{\text{min}}$ suppression. 

The exponential term extracted from the fit is considered as the maximum possible cross section of DFR since it may include other contributions to the low-$|t|$ enhancement in the Kopeliovich prediction in addition to DFR. 
It shows a slightly softer shape in $|t|-|t|_{\text{min}}$ than the Rein model prediction (Fig. \ref{figure_dfr_tmtmin}, right). 
An integral over the exponential 
gives the estimation of the total DFR cross section at $E_\nu=2.7$\,GeV: $3.04\times10^{-40}\,\text{cm}^{2}/\text{proton}$. 
For the measurement reported in this paper, the DFR background events are first simulated by the Rein model in GENIE 2.12.2, and then reweighted to the estimation from Kopeliovich in both normalization and shape as a function of $|t|-|t|_{\text{min}}$. 
This reweighting makes a 1\% difference in the coherent signal measurement. 

The above method simulates DFR independently from other pion production channels on hydrogen simulated by GENIE. 
The interference between DFR and non-DFR channels, however, could potentially affect the rate and shape of both DFR and non-DFR pion productions, and the effect on the measurement reported in the paper needs to be discussed. 
Since the Kopeliovich model includes both DFR and non-DFR contributions in a coherent way, 
this effect can be taken into account by simulating all the $\nu p \rightarrow \nu p \pi^0$ events on hydrogen using the Kopeliovich model. 
This is achieved by reweighting the GENIE simulated $\nu p \rightarrow \nu p \pi^0$ events on hydrogen to the prediction of Kopeliovich as a 2D function of $|t|$ and $Q^2$.
The background template fit is repeated with the hydrogen contribution fixed.  
The effect on the measurement is a 2.6\% difference from the nominal result, which is considered as an additional systematic uncertainty contribution from DFR. 


\begin{thebibliography}{46}
\bibitem{Adler:1964pcac} S. Adler, Phys. Rev. \textbf{135B}, 963 (1964).

\bibitem{Rein:1983rs} D. Rein and L. M. Sehgal, Nucl. Phys. \textbf{B223}, 29 (1983).

\bibitem{Rein:2007rs} D. Rein and L. M. Sehgal, Phys. Lett. B \textbf{657}, 207 (2007). 

\bibitem{Andreopoulos:2010genie} C. Andreopoulos \textit{et al.}, Nucl. Instrum. Methods Phys. Res. Sect. A \textbf{614}, 87 (2010); C. Andreopoulos \textit{et al.}, arXiv:1510.05494v1 [hep-ph] (2015).

\bibitem{Berger:2009bs} C. Berger and L. M. Sehgal, Phys. Rev. D \textbf{79}, 053003 (2009).

\bibitem{Belkov:1987bk} A. A. Belkov and B. Z. Kopeliovich, Sov. J. Nucl. Phys. \textbf{46}, 499 (1987).

\bibitem{Kopeliovich:2005bzh} B. Z. Kopeliovich, Nucl. Proc. Suppl. \textbf{139}, 219 (2005).

\bibitem{Hernandez:2009eh} E. Hernandez, J. Nieves, and M. J. Vicente Vacas, Phys. Rev. D \textbf{80}, 013003 (2009).


\bibitem{Kartavtsev:2006eap} A. Kartavtsev, E. A. Paschos, and G. J. Gounaris, Phys. Rev. D \textbf{74}, 054007 (2006). 


\bibitem{Paschos:2009eap} E. A. Paschos, D. Schalla, Phys. Rev. D \textbf{80}, 033005 (2009). 

\bibitem{Singh:2006microscopic1} S. K. Singh, M. S. Athar, and S. Ahmad, Phys. Rev. Lett. \textbf{96}, 241801 (2006).

\bibitem{Alvarez-Ruso:2007microscopic2} L. Alvarez-Ruso, L. S. Geng, S. Hirenzaki, and M. J. Vicente Vacas, Phys. Rev. C \textbf{75}, 055501 (2007). 

\bibitem{Alvarez-Ruso:2007microscopic2nc} L. Alvarez-Ruso, L. S. Geng, and M. J. Vicente Vacas, Phys. Rev. C \textbf{76}, 068501 (2007). 

\bibitem{Amaro:2009microscopic3} J. E. Amaro, E. Hernandez, J. Nieves, and M. Valverde, Phys. Rev. D \textbf{79}, 013002 (2009). 

\bibitem{Leitner:2009microscopic4} T. Leitner, U. Mosel, and S. Winkelmann, Phys. Rev. C \textbf{79}, 057601 (2009). 

\bibitem{Hernandez:2009microscopic5} E. Hernandez, J. Nieves, and M. J. Vicente Vacas, Phys. Rev. D \textbf{80}, 013003 (2009). 

\bibitem{Nakamura:2009microscopic6} S. X. Nakamura, T. Sato, T.-S. H. Lee, B. Szczerbinska, and K. Kubodera, Phys. Rev. C \textbf{81}, 035502 (2010).

\bibitem{Faissner:1983ap} H. Faissner \textit{et al.} (Aachen-Padova Collaboration),  Phys. Lett. B \textbf{125}, 230 (1983).

\bibitem{Isiksal:1984Gargamelle} E. Isiksal, D. Rein, and J. G. Morfin (Gargamelle Collaboration), Phys. Rev. Lett. \textbf{52}, 1096 (1984).

\bibitem{Baltay:1986bc} C. Baltay {\it et al.} (Columbia-BNL Collaboration), Phys. Rev. Lett. \textbf{57}, 2629 (1986).

\bibitem{Bergsma:1985charm} F. Bergsma \textit{et al.} (CHARM Collaboration), Phys. Lett. \textbf{157B}, 469 (1985).

\bibitem{Grabosch:1986SKAT} H. J. Grabosch \textit{et al.} (SKAT Collaboration), Z. Phys. C \textbf{31}, 203 (1986).

\bibitem{Aguilar-Arevalo:2008miniboone} A. A. Aguilar-Arevalo \textit{et al}. (MiniBooNE Collaboration), Phys. Lett. B \textbf{664}, 41 (2008). 

\bibitem{Kurimoto:2008sciboone} Y. Kurimoto \textit{et al}. (SciBooNE Collaboration), Phys. Rev. D \textbf{81}, 111102 (2010). 

\bibitem{Kullenberg:2009nomad} C. T. Kullenberg \textit{et al.} (NOMAD Collaboration), Phys. Lett. B \textbf{682}, 177 (2009).

\bibitem{Adamson:2016minos} P. Adamson {\it et al.} (MINOS Collaboration), Phys. Rev. D {\bf 94}, 072006 (2016).

\bibitem{Ayres:2007nova} D. S. Ayres \textit{et al.}, NOvA Technical Design Report, No. FERMILAB-DESIGN-2007-01, 2007.

\bibitem{Adamson:2016numi} P. Adamson \textit{et al.}, Nucl. Instrum. Methods Phys. Res. Sect. A \textbf{806} 279 (2016), FERMILAB-PUB-15-253-AD-FESS-ND.

\bibitem{Hamamatsu:apd} The NOvA APD is a custom variant of the Hamamatsu S8550, http://www.hamamatsu.com/us/en/product/ alpha/S/4112/S8550-02/index.html. 

\bibitem{Bohlen:2014fluka} T. T. Bohlen \textit{et al.}, Nucl. Data Sheets \textbf{120}, 211 (2014); A. Ferrari \textit{et al.}, CERN Reports No. CERN-2005-10, No. INFN/TC 05/11, and No. SLAC-R-773, 2005.

\bibitem{Campanella:1999flugg} M. Campanella \textit{et al.}, CERN Technical Report No. CERN-ATL-SOFT-99-004, 1999.

\bibitem{Agostinelli:2003geant4} S. Agostinelli \textit{et al.}, Nucl. Instrum. Methods Phys. Res. Sect. A \textbf{506}, 250 (2003); J. Allison \textit{et al.}, IEEE Trans. Nucl. Sci. \textbf{53}, 270 (2006).

\bibitem{Aliaga:2016flux} L. Aliaga  {\it et al.} (MINERvA Collaboration), Phys. Rev. D \textbf{94}, 092005 (2016).

\bibitem{Rein:1981resRS} D. Rein and L. M. Sehgal, Ann. Phys. (N.Y.) 133, 79 (1981).

\bibitem{Bodek:2005by} A. Bodek, I. Park, and U.-K. Yang, Nucl. Phys. B. Proc. Suppl. \textbf{139}, 113 (2005). 

\bibitem{Rein:1986dfr} D. Rein, Nucl. Phys. \textbf{B278}, 61 (1986). 

\bibitem{Kopeliovich:2012} B.Z. Kopeliovich, I. Schmidt, and M. Siddikov, Phys. Rev. D \textbf{85}, 073003 (2012).
\bibitem{Kopeliovich:np} B.Z. Kopeliovich \textit{et al.}, Neutrinoproduction of pions off nuclei, http://atlas.fis.utfsm.cl/np/.

\bibitem{Bodek:1981br1} A. Bodek and J. L. Ritchie, Phys. Rev. D \textbf{23}, 1070 (1981). 
\bibitem{Bodek:1981br2} A. Bodek and J. L. Ritchie, Phys. Rev. D \textbf{24}, 1400 (1981).


\bibitem{Fernandes:2008hough} L. Fernandes and M. Oliveira, Patt. Rec. \textbf{41}, 299 (2008).

\bibitem{Niner:2015thesis} E. Niner, Ph.D. thesis, Indiana University, 2015， FERMILAB-THESIS-2015-16.  
\bibitem{Sachev:2016thesis} K. Sachdev,  Ph.D. thesis, University of Minnesota, 2015, FERMILAB-THESIS-2015-20.  

\bibitem{Adamson:2017zcg} 
  P.~Adamson {\it et al.} (NOvA Collaboration),
  Phys.\ Rev.\ D {\bf 96}, 072006 (2017).
  
\bibitem{Tanabashi:pdg2018} M. Tanabashi  \textit{et al}. (Particle Data Group), Phys. Rev. D, \textbf{98}, 030001 (2018).

\bibitem{Higuera:2014minerva}  A. Higuera \textit{et al.} (MINERvA Collaboration) Phys. Rev. Lett. \textbf{113}, 261802 (2014).

\bibitem{Mislivec:2017qfz} 
  A.~Mislivec {\it et al.} (MINERvA Collaboration),
  Phys.\ Rev.\ D {\bf 97}, 032014 (2018).



\bibitem{Duyang:2015mrbrem} 
  H. Duyang (NOvA Collaboration),
  arXiv:1511.00351.

\end{thebibliography}
\end{document}